\documentstyle[prb,preprint,aps,eqsecnum,tighten,epsf,rotate]{revtex}
\def\plotfig#1#2{\rotate[#1]{\epsfysize=\textwidth \epsfbox[0 0 595
                  800]{#2}}}
\begin{document}

\title{Elastic theory of flux lattices in presence of weak disorder}
\author{Thierry Giamarchi}
\address{Laboratoire de Physique des Solides, Universit{\'e} Paris--Sud,
                   B{\^a}t. 510, 91405 Orsay, France\cite{junk}}
\author{Pierre Le Doussal}
\address{Laboratoire de Physique Th{\'e}orique
de l'Ecole Normale Sup{\'e}rieure \cite{frad}, 24 Rue Lhomond, F-75231 Paris
Cedex,
France}
\maketitle

\begin{abstract}
The effect of weak impurity disorder on
flux lattices at equilibrium is studied quantitatively
in the absence of free dislocations using
both the Gaussian variational method and the renormalization group.
Our results for the mean square relative displacements $\tilde{B}(x)=
\overline{\langle u(x)-u(0) \rangle ^2}$ clarify
the nature of the crossovers with distance.
We find three regimes: (i) a short distance
regime (``Larkin regime'') where elasticity holds (ii) an intermediate
regime (``Random Manifold'') where vortices are pinned independently (iii)
a large distance, quasi-ordered regime where the
periodicity of the lattice becomes important. In the last regime
we find universal logarithmic growth of displacements for $2<d<4$:
$\tilde{B}(x) \sim A_d \log|x|$
and persistence of algebraic quasi-long range translational order.
The functional renormalization group to $O(\epsilon=4-d)$ and the
variational method, when they can be compared, agree within $10\%$ on
the value of $A_d$. In $d=3$  we compute the function
describing the crossover between the
three regimes. We discuss the observable signature
of this crossover in decoration experiments
and in neutron diffraction experiments on flux
lattices. Qualitative arguments are given suggesting the
existence for weak disorder in $d=3$ of a `` Bragg glass ''
phase without free dislocations and with algebraically
divergent Bragg peaks.
In $d=1+1$ both the variational method and the Cardy-Ostlund
renormalization group predict a glassy state below
the same transition temperature $T=T_c$,
but with different $\tilde{B}(x)$ behaviors.
Applications to $d=2+0$ systems and experiments
on magnetic bubbles are discussed.

\end{abstract}
\pacs{74.60.Ge, 05.20.-y}
\narrowtext

\section{Introduction}

The interest in the pinning of the Abrikosov vortex
lattice by impurities was revived recently with
the discovery of high-Tc superconductors.
Impurity disorder conflicts with the long range
translational order of the flux lattice and
some glassy state is generally expected to appear.
Understanding the precise nature of this new thermodynamic state
and how it depends on the type of disorder existing
in the system is very important for the determination
of the transport properties
of these materials, such as critical currents and I-V characteristics
\cite{feigelman_collective,fisher_vortexglass_short,%
nelson_columnar_long,hwa_splay_prl,revue_russes}.
This problem, however, is only one aspect of
the more fundamental and broader question of the effect of quenched impurities
on
any translationally ordered structure, such as a crystal.
This question arises in a large number of physical systems
under current active experimental study.
Examples are charge density waves \cite{gruner_revue_cdw},
Wigner crystals
\cite{wigner_andrei,wigner_jiang,wigner_shklovskii,wigner_millis}
magnetic bubbles \cite{seshadri_bubbles_thermal,seshadri_bubbles_long},
Josephson junctions \cite{vinokur_josephson_short,balents_josephson_long},
the surface of crystals with quenched bulk or substrate disorder
\cite{toner_log_2}, domain walls in incommensurate solids
\cite{pokrovsky_talapov_prl}.
All these systems have in common a perfectly ordered
underlying structure modified by elastic distortions and
possibly by topological defects such as dislocations, due to
temperature or disorder.
The effect of thermal fluctuations alone on
three-dimensional and especially on two-dimensional structures
is by now well understood, and it was shown that
topological defects are not important in the low temperature
solid phase \cite{nelson_halperin_melting}.
Much less is known
however on the additional effects of quenched disorder. In particular,
the important question of precisely how quenched disorder destroys
the translational long range order
of the lattice is far from being elucidated.
If disorder is strong, the underlying order
is a priori destroyed at every scale and an analytical
description of the problem starting from the Abrikosov lattice
is difficult. One then has to use a more macroscopic
approach based on phenomenological models such as
the gauge glass models
\cite{fisher_vortexglass_short,fisher_vortexglass_long,bokil_young_vglass}.
The success of these approaches
then crucially depends on whether these effective models are
indeed a good representation of the system at large scale,
a largely uncontrolled assumption.
If disorder is weak enough, however, one expects the perfect lattice
to survive at short scales. Thus a natural first step for a theoretical
description is to neglect dislocations and to treat
the simpler problem of an elastic medium submitted to weak
impurities. In that case one can consider a gaussian random
potential created by many weak impurities with short-range
correlations. In this paper we will focus on point-like disorder.
The problem of correlated disorder, such as
columnar or twin boundary pinning in superconductors
\cite{nelson_columnar_long,hwa_splay_prl}, which is also relevant
for any type of {\it quantum}
problems can be treated by similar methods, and is
examined in details in Ref. \onlinecite{giamarchi_columnar_variat}.

This simpler problem of an elastic lattice in presence
of weak disorder is already quite non trivial. Despite
several attempts, its physics has not been completely understood.
An important quantity, which measures
how fast translational order decays, is the translational
correlation function $C_{K_0}(r)=
\overline{\langle e^{i K_0 \cdot (u(r)-u(0))} \rangle}$,
where $u(r)$ is the displacement from the perfect lattice and
$K_0$ one of the first reciprocal lattice vector. We denote by
$\langle \rangle$ and $\overline{\text{\ \ \ }}$, the thermodynamic
average and the disorder average respectively.
$C_{K_0}(r)$ can be
extracted from the Fourier transform of the density-density correlation
function at wavevectors near $q=K_0$, or directly measured by imaging the
deformed lattice, and is thus a quantity easily accessible in experiments.

A first calculation of $C_{K_0}(r)$ was performed
by Larkin \cite{larkin_70} using a model
in which weak {\it random forces} act independently on each vortex.
These forces are correlated over a small length $\xi_0$,
of the order of the superconducting
coherence length. This model predicts that weak disorder destroys translational
 order below
four dimensions \cite{larkin_70} and $C_{K_0}(r) \sim
\text{exp}(-r^{4-d})$.
The destruction of the long range order
in this simple gaussian model can be understood
easily from the standard Imry-Ma argument. To accommodate the random
forces a region of size $R$ will undergo relative deformations of order $u$.
The cost in elastic energy is $R^{d-2} u^2$, while the gain
in potential energy is $u R^{d/2}$. The optimal is thus $u=R^{(4-d)/2}$
leading to the above decay of $C_{K_0}(r)$. Using similar arguments
in the presence of an external Lorentz force, Larkin and Ovchinikov
\cite{larkin_ovchinnikov_pinning}
constructed a theory of collective pinning of the flux lattice.
In this theory the critical current is determined
from the length scale $R_c$ at which relative
displacements are or order $u \approx \xi_0$. This theory was
very successful in describing conventional superconductors.
However, a need to reconsider this theory was prompted by high Tc
superconductors
where the flux lattice is usually probed at larger scales.
It turns out that the Larkin model, while it is useful for estimating
critical currents, cannot be used to study large scale quantities
such as translational order.

In fact the purely gaussian model with random forces,
and the resulting linear elasticity, becomes inadequate
beyond the Larkin-Ovchinikov length $R_c$. It has only one trivial equilibrium
state
and responds linearly to external force. It is thus
too simple to approximate correctly the full non linear problem
and grossly overestimates the effect of disorder. At larger
scales the lattice starts behaving collectively as an elastic
manifold in a {\it random potential} with many metastable states,
thus the exponential
decay of $C_{K_0}(r)$
in $d=3$ cannot hold beyond $r>R_c$.
Using known results on the so-called ``random manifold'' problem,
Feigelman et al. \cite{feigelman_collective} showed that
the system presents glassy behavior
and computed transport properties.
This was also pointed out by Bouchaud, M\'ezard
and Yedidia (BMY)
\cite{bouchaud_variational_vortex,bouchaud_variational_vortex_prl},
who used the Gaussian Variational Method (GMV) to study this problem,
and found a
power-law roughening of the lattice with stretched exponential decay
of $C_{K_0}(r)$.

However, the {\it periodicity} of the lattice was not
properly taken into account in all the above works.
The periodicity has
important consequences for the behavior of correlation functions
at large scales.
Indeed, it was suggested with the use of qualitative Flory arguments,
that
periodicity leads to logarithmic roughening \cite{nattermann_pinning},
rather than a power law.

In this paper we develop a quantitative description of the static
properties of a lattice in the presence of disorder. A short
account of some of the results of this paper were presented in a recent
letter \cite{giamarchi_vortex_short}. We take into account
both the existence of many metastable
states and the periodicity of the lattice. One of the difficulties in
the physics of this problem is that the disorder
varies at a much shorter length scale than the lattice spacing. As a
consequence the elastic limit has to be taken with some care. Indeed
in this limit the displacement varies slowly, but the density still
consists in a series of peaks. To couple the density to the random potential
it is thus important to distinguish between its slowly varying components
$\Delta_{q=0}$ and its Fourier component $\Delta_{K_0}$
close to the periodicity of the lattice. This separation of harmonics exposes
clearly the physics and allows to treat all the regimes in length scales
in a simple way.
To study the resulting model, we mainly use the Gaussian Variational
Method, developed to study manifold in random media
\cite{mezard_variational_replica}. We also use the Renormalization Group
(RG) close to $d=4$ dimensions and in $d=2$ dimensions.
Comparison of the two methods
provides a confirmation of the accuracy of the GVM.

One the main results of the present study, which is
somewhat surprising in view of conventional wisdom based on
Larkin's original calculation, is that quasi-long range order
survives in the system.
This means that $C_{K_0}(r)$ decays as a power law at large distance. Such a
property for disordered lattice is similar to the quasi-order found
for clean two-dimensional solids. This state however has the peculiar property
of being a glass with many metastable states,
and at the same time show Bragg peaks as a solid does. For these
reasons we would like to call it a ``Bragg glass''. Note that
this is a much stronger property that the so-called
``hexatic glass'' \cite{chudnovsky_pinning,grier_decoration_manips}
since hexatic order in the elastic limit
is a straightforward consequence of the absence of dislocations.
In the Bragg glass,
two important length scales control the crossover towards the
asymptotic decay, a consequence of the
fact that the disorder varies at a much smaller scale than the lattice
spacing $a$. When the mean square of the relative displacement
$\overline{\langle [u(x)-u(0)]^2 \rangle}$ of
two vortices as a function of their separation $x$ is shorter than the
square of the Lindemann length $l_T^2 = \langle u^2 \rangle$, the
thermal wandering of the
lines averages enough over the random potential and the model becomes
equivalent to the random force Larkin model. At low enough
temperature, $l_T$ is replaced by the correlation length of the random
potential
$\xi_0$, which is of the order of the superconducting coherence length. In that
case the
crossover length is $R_c$ \cite{feigelman_collective,%
bouchaud_variational_vortex_prl,bouchaud_variational_vortex}. When the
relative displacement is larger than the correlation length of the
random potential but smaller than the lattice spacing $a$,
this is the random manifold regime where each line sees effectively
an independent random potential. When the
relative displacement becomes larger than the lattice spacing, one enters the
asymptotic quasi-ordered regime. This occurs for separations of order
$\xi$. In general the two lengths $R_c$ and $\xi$ are widely different.
The theory developed here can be applied to any elastic system in
the elastic limit $\xi \gg a$. In relation with experimental
systems we focus particularly on the triangular Abrikosov lattice in
$d=2+1$,
point vortices in thin films and magnetic bubbles in $d=2+0$, and
will also mention lines in a plane $d=1+1$.

In this paper we will not treat topological defects quantitatively. Although
a full description of a lattice in presence of disorder
should also include topological defects, their effect might
not be as severe as
commonly believed from misleading Imry-Ma type arguments.
Indeed the fact that within the elastic theory quasi-long range
order is preserved
at large distances makes the system much more stable to dislocations.
Since in $d=3$ the core energy of a dislocation increases as
its size $L$, it is actually very possible that a phase
without unbound dislocations exists
in $d=3$ in the presence of weak disorder.
Indeed,
Bitter decoration experiments \cite{grier_decoration_manips} at the
highest fields available, about 70 G for these low-fields experiments,
show remarkably large regions free of dislocations.
In recent neutron experiments \cite{yaron_neutrons_vortex}
it was shown that the degree
of order depends on the way the system is prepared.
A more perfect lattice with a smaller number of dislocations
was prepared by first driving the system at a velocity
high enough for translational order to heal,
and then slowing it back down to zero velocity. It is thus
conceivable that in $d=3$ the presence of dislocations is
overall a non-equilibrium feature.
In two dimensions ($d=2+0$), dislocations
are energetically less costly and will probably appear at large scales,
although this has not yet been firmly established. However,
as we will discuss here, the length
scale between unbound dislocations $\xi_D$ can be much larger than $\xi$
in a low temperature regime. In that regime the
main cause of the decay of translational order is elastic deformations
due to disorder.

The paper is organized as follows:
For convenience we have separated
the mostly technical sections (\ref{simplified}) and
(\ref{degal4}) from the one discussing applications to
physical systems (\ref{degal3}) and (\ref{degal2}).
In Section (\ref{derivation}), we introduce the model and
derive the correct elastic limit. Simple dimensional arguments
\`a la Fukuyama-Lee are given to identify the relevant length
scales.
In section (\ref{simplified}) we apply the
Gaussian Variational Method to a simplified isotropic
version of the model.
Thus the method can be exposed
without being obscured by
unnecessary technical complications specific to real vortex lattices
such as anisotropy and non-local elasticity, while
the essential physics is retained.
This section contains most of the technical
details and methods used. We explain why a previous
application of the variational method by BMY led to erroneous conclusions
about the behavior at large scales.
In Section (\ref{degal3}) we apply the theory to
$d=2+1$ solids such as the vortex lattices
using a realistic elastic Hamiltonian. We compute the
translational order correlation function $C_{K_0}(r)$
with the full crossover between the three regimes in
distance. We discuss the experimental signatures
for decoration and neutron diffraction experiments.
In particular the results of a comparison
between decoration images and theoretical predictions
are mentioned and detailed predictions are made
for the neutron diffraction intensities.
We then give a simple physical interpretation of the
various regimes in distance and also
argue that dislocations are less likely to appear as
commonly believed.
In Section (\ref{degal4}) we apply the functional renormalization group in
$d=4-\epsilon$, and compare its findings with those of the variational
method.
In Section (\ref{degal2}) we examine two-dimensional
systems, for which thermal fluctuations play a
more important role. We first apply the variational
method which shows that below a critical temperature
$T_c$ the system is glassy with logarithmic growth
of displacements. This is compared to predictions of the
renormalization group in $d=1+1$.
We then give a physical discussion of what should be
expected for $d=2+0$ systems, such as magnetic bubbles,
where dislocations have to be considered.
Conclusion can be found in Section \ref{conclusion}.
Finally the bold and brave can look at the appendices
where most of the most tedious technicalities are relegated.

\newpage

\section{Derivation of the model and physical content} \label{derivation}

\subsection{A General elastic Hamiltonian}

In the absence of disorder the vortices form, at equilibrium, a perfect
lattice of spacing $a$ whose sites are labeled by an integer
$i$ and position will be denoted by $R_i$. Since we want to apply this
theory also to a lattice of vortex lines, we consider the more general case
where
the $R_i$ are $m$-component
lattice vectors and there is in addition $d-m$ transverse directions
denoted by $z$
so that the total spatial dimension is $d$. Throughout this paper we will
denote the $d$
dimensional coordinates by $x\equiv(r,z)$ and
similarly the Fourier space (momentum) coordinates by $q\equiv(q_\perp,q_z)$.
For example, the Abrikosov
lattice corresponds to $m=2$ and $d=3$ and $z$ is along the direction
of the magnetic field. The displacements
relative to the equilibrium positions are denoted by the $m$ component
vector $u(R_i,z)\equiv u_i(z)$.
For weak disorder ($a/\xi \ll 1$ where $\xi$ is defined below ) and in the
absence of dislocations,
it is legitimate to assume that $u(R_i,z)$ is slowly
varying on the scale of the lattice and to use a continuum elastic energy,
as a function of the continuous variable $u(x)$. We consider the simple
elastic Hamiltonian:
\begin{equation} \label{real}
H_{\text{el}} = \frac12 \sum_{\alpha,\beta}
\int_{\text{BZ}} \frac{d^dq}{(2\pi)^d}
u_\alpha(q) \Phi_{\alpha\beta}(q) u_\beta(-q)
\end{equation}
where $\alpha,\beta=1,..m$ labels the coordinates, and $\text{BZ}$ denotes the
Brillouin zone.
The $\Phi_{\alpha\beta}$ are the elastic matrix.
Such an elastic description is valid as
long as the {\it relative} displacement of two neighboring points remains
small i.e. $|u(R_i)-u(R_{i+1})| \ll a$, but does {\it not} suppose the
individual displacements themselves to be small.
We differ the study of the realistic elastic Hamiltonian (\ref{real})
until sections \ref{degal3} and \ref{degal2}, and in order to illustrate
the method in sections \ref{derivation} and \ref{simplified}
we use the fully isotropic elastic Hamiltonian
\begin{equation} \label{iso}
H_{\text{el}} = \frac{c}2 \int d^dx (\nabla u(x))^2
\end{equation}
corresponding to the case $\Phi_{\alpha\beta}(q) = c q^2 \delta_{\alpha
\beta}$.

In the limit where many weak impurities act collectively on a vortex
the disorder can be modeled by
a gaussian random potential $V(x)$ with correlations:
$\overline{V(x)V(x')}=\Delta(r-r')\delta(z-z')$ where $\Delta(r)$
is a short range function of range $\xi_0$ (the superconducting
coherence length) and Fourier transform $\Delta_{q_\perp}$.
The other limit, corresponding to a few strong pins,
can be modeled by a Poissonian distribution and will not be
considered here.
Since the density of vortices at a given point is given by
\begin{equation} \label{densistart}
\rho(x) = \sum_i \delta(r - R_i -u(R_i,z))
\end{equation}
the total Hamiltonian is therefore
\begin{equation} \label{total}
H = H_{\text{el}} + \int d^dx V(x) \rho(x)
\end{equation}

The simplest way to recover translational invariance, is to use
the well-known replica trick \cite{edwards_replica}.
This amounts to introduce $n$ identical
systems by replicating the original Hamiltonian. It is then
possible to average over disorder, the proper quenched average being
recovered in the limit $n\to 0$.
After replicating, the interaction term becomes
\begin{equation} \label{intor}
H_{\text{pin}} = - \frac1{2T} \sum_{a,b} \int d^dx d^dx'
            \Delta(r-r') \delta(z-z') \rho^a(x) \rho^b(x')
\end{equation}
where $a,b$ are the replica indices. (\ref{intor})
can be rewritten
\begin{eqnarray} \label{interaction}
H_{\text{pin}} & = & - \frac1{2T} \sum_{a,b} \int d^dx d^dx' \sum_{i,j}
            \Delta(x-x')\delta(z-z')
            \delta(x - R_i - u^a_i(z)) \delta(x - R_j - u^b_j(z))
            \nonumber \\
            & = & - \frac1{2T} \sum_{a,b} \sum_{i,j} \int d^{d-m}z
                  \Delta(R_i - R_j + u^a_i(z) - u^b_j(z))
\end{eqnarray}
Where $a,b=1,\ldots,n$ are the replica indices. As we show in the
following section it is extremely
important to keep the discrete nature of the lattice in
(\ref{interaction}), and the continuum limit of $H_{\text{pin}}$
should be done with
some care.

\subsection{Decomposition of the density}

Using the form (\ref{interaction}) for the Hamiltonian in term of
the displacement fields $u_i$ is rather cumbersome.
(\ref{interaction}) leads to
a non local theory,
even in the limit where the disorder is
completely uncorrelated $\Delta(r-r')=\delta(r-r')$. Indeed, vortices
belonging to two different replica sets
can be a priori at the same point in space $r = R_i + u^a_i(z) =
R_j + u^b_j(z)$
while having very different equilibrium positions $R_i\ne R_j$.
This occurs when the displacements of the vortices are
large enough. Since $R_i$,
the equilibrium position of the vortices, have clearly no
physical significance, except as an internal label,
it is much more convenient to use instead a label that is a function
of the actual position of the vortices.
This can be achieved by introducing the
slowly varying field
\begin{equation} \label{slowly}
\phi(r,z) = r - u(\phi(r,z),z)
\end{equation}
Such a field will allow the continuum limit of
(\ref{interaction}) to be taken easily.
For each configuration
of disorder, or alternatively for each replica set, one introduces
a different field $\phi$. Such a labeling
is always exact when the transverse dimension $m$ is $m=1$
\cite{haldane_bosons}. In
more than one transverse dimension, this representation assumes the absence of
dislocations in the system. In a self consistent manner, we will justify
a posteriori both assumptions of elasticity and absence of dislocations
using the solution in sections (\ref{selfconsist}) and (\ref{discussion}).

Using $\phi(x)$ the density can be rewritten as (see
appendix~\ref{density})
\begin{equation} \label{transp}
\rho(x) =  \rho_0 \text{det}[\partial_\alpha \phi_\beta] \sum_K
e^{i{K}\cdot\phi(x)}
\end{equation}
where the $K$ are the vectors of the reciprocal lattice and $\rho_0$
is the average density.
In the  elastic limit one can expand (\ref{transp}) to get
\begin{equation} \label{transparent}
\rho(x) \simeq \rho_0 (1 - \partial_\alpha u_\alpha (\phi(x)) +
        \sum_{K \ne 0} e^{i K r} \rho_K(x))
\end{equation}
where
\begin{equation}
\rho_K(x)=e^{-i K \cdot u(\phi(x))}
\end{equation}
is the usual translational order parameter
defined in terms of the reciprocal lattice vectors $K$.
Expression (\ref{transparent}) respects the periodicity of the lattice
i.e. is obviously invariant by a global translation $u \to u +a$.
Another advantage of the decomposition (\ref{transparent}) is
that the various Fourier components of the density relative to the
periodicity of the ordered lattice appear clearly in $H_{\text{pin}}$
\begin{equation} \label{couplage}
\int dx V(x) \rho(x) = -\rho_0 \int dx V(x)  \partial_\alpha u_\alpha
 + \rho_0 \int dx \sum_{K\ne 0} V_{-K}(x) \rho_K(x)
\end{equation}
where
\begin{equation}
V_K(x) = V(x) e^{- i K r}
\end{equation}
is the part of the random potential with Fourier components close to $K$.
Since the
energy is invariant when changing $u\to u + a$, $u$
itself cannot appear in the expression (\ref{couplage}), but  only
$\partial u$, and in principle
higher derivatives, or a periodic function of $u$ are allowed.

The first term in the right hand side of (\ref{couplage})
is the part of the deformation of the lattice
that couples to the long wavelength of the disorder potential. It
results in an increase or decrease of the average density in regions
where the potential is favorable or unfavorable.
The second term couples to the higher Fourier
components of the disorder. The average density is not affected but the
lattice can be shifted so that the lines sit in the minimum of the
disorder potential.
In the usual elasticity theory, one takes the continuum limit
for the displacement field {\bf and} assumes that the density itself
is smooth on the scale of the lattice. This allows to keep only the
gradient term in (\ref{couplage}).
Here although it is possible to take the continuum
limit for the displacements $u$ since they vary slowly on the scale of
the vortex lattice ( $\nabla u \ll 1$) , it is imperative to retain the
discrete nature of
the density. This is because the scale at which the disorder varies (for
superconductors
it is comparable to the scale of the real atomic crystal) is usually shorter
than the lattice spacing of the vortex lattice itself.

If one uses the representation (\ref{transparent}) of the density and
(\ref{interaction}) and discard spatial averages of rapidly oscillating terms,
the replicated Hamiltonian becomes, in the isotropic case:
\begin{eqnarray} \label{cardyos}
H_{\text{eff}} & = & \frac{c}2 \int d^dx (\nabla u(x))^2 \\
& & - \int d^dx  \sum_{a,b}
[\frac{\rho_0^2\Delta_0}{2T}\partial_\alpha
 u_\alpha^a\partial_\beta u_\beta^b
+ \sum_{K\ne 0} \frac{\rho_0^2\Delta_K}{2T}
\cos(K \cdot (u^{a}(x)-u^{b}(x)))]
\nonumber
\end{eqnarray}
\narrowtext%
To be rigorous the last terms in (\ref{cardyos}) should be written in term of
$u(\phi(x))$ rather than $u(x)$, but this has no effect on our results. It
leads only to
corrections of higher order in $\nabla  u$ which we neglect since
we work in the elastic limit $a/\xi \ll 1$.
The Hamiltonian (\ref{cardyos}) will be our starting model, and from now
on we absorb the coefficient $\rho_0^2$ in $\Delta_K$,
$\rho_0^2 \Delta_K \to \Delta_K$.

A general property of the Hamiltonian (\ref{cardyos}) is the invariance
of the disorder term under the transformation $u_a(x) \to u_a(x) + w(x)$
where $w(x)$ is an arbitrary function of $x$. This statistical invariance {\it
guarantees} that
the elastic term in (\ref{cardyos}) is unrenormalized by disorder. Note that
in the {\it original} non local model (\ref{interaction}) this symmetry is only
approximate,
and indeed one would find there a small (of order $(a/\xi)^2$ ) and
unimportant renormalization of the elastic coefficients by disorder.

The principal quantities of interest are the mean squared relative displacement
$\tilde{B}(x)$ of two vortices, averaged over disorder, which
is determined by the correlation of $u$ diagonal in replicas
\begin{eqnarray} \label{tilde} & &
\tilde{B}(x) = \frac1m \overline{\langle (u(x) - u(0))^2 \rangle } =
2 T \int {d^dq \over (2\pi)^d}
(1 - \cos(qx) ) \tilde{G}(q) \\ & &
\langle u^a_\alpha(q) u^a_\beta(-q) \rangle = \delta_{\alpha \beta}
T \tilde{G}(q)            \nonumber
\end{eqnarray}
and the translational order
correlation function $C_K(x)$
\begin{equation}
C_K(x) = \overline{\langle \rho^*_K(x) \rho_K(0) \rangle}
\end{equation}
In the gaussian theory that will be considered below,
the two correlation functions are simply related by
\begin{equation}
C_K(x) = e^{- \frac{K^2}2 \tilde{B}(x)}
\end{equation}

The Hamiltonian (\ref{cardyos}) can be applied directly to study quantum models
with a time dependent disorder. A more physical disorder for quantum
systems would be only
space dependent. This would correspond to correlated disorder
in one (the ``time'') direction for the classical system and can be
studied by the same method than the one used in this paper
\cite{giamarchi_columnar_variat}. For completeness
we also give here in appendix~\ref{quantuma}
the connection between the quantum mechanics
of interacting bosons and fermions and an elastic system in $d=1+1$
dimensions.

\subsection{Dimensional Arguments}
\label{dimensionarg}

Before starting the full calculation,
let us estimate the effects of the different terms in
(\ref{couplage}) in a way similar to Ref. \onlinecite{fukuyama_pinning}.
In the presence of many weak pins, $u$ cannot
distort to take advantage of each of them, due to the cost in elastic
energy. One can assume that $u$
varies of $\sim a$ over a length $\xi \gg a$. The density of kinetic
energy is $\sim c(a/\xi)^2$, where $c$ is an elastic constant.
The various Fourier components of the
disorder will give different contributions. The long
wavelength part of the disorder gives
\begin{equation}
H^{\text{dis}}_{q \sim 0} \sim \rho_0 \int d^dx V(x) \partial_\alpha
u_\alpha(x)\sim \Delta_0^{1/2} a /\xi^{1+d/2}
\end{equation}
For the higher Fourier components the disorder term
\begin{equation} \label{disorder}
H^{\text{dis}}_{q \sim K}
 = \rho_0 \int d^dx V(x) e^{i K x} e^{-i K u(x)}
\end{equation}
can be estimated over the volume $\xi^d$ as
$e^{-i K u} \int_{\xi^d} d^dx V(x) e^{i K x}$.
This sum can be viewed as a random walk in the complex plane
\cite{fukuyama_pinning} and
the value of $u$ adjusts itself to match the
phase of the random potential. Therefore the gain in
energy density due to the disorder term is of order
\begin{equation}
H^{\text{dis}}_{q \sim K} \sim \Delta_K^{1/2} /\xi^{d/2}
\end{equation}
Optimizing the gain in potential energy gain versus the cost in elastic energy
determines $\xi$. One can therefore associate to each Fourier component a
length scale
above which the corresponding disorder will be relevant and destroy the perfect
lattice
\begin{eqnarray} \label{leslong}
\xi_{q \sim 0} & \sim & a\left(c^2 a^{d}/\Delta_0\right)^{1/(2-d)} \qquad
\text{if $d<2$} \\
\xi_{q \sim K} & \sim & a\left(c^2 a^{d}/\Delta_K\right)^{1/(4-d)}
\end{eqnarray}
The $q\sim 0$ component of the disorder is relevant only for
$d \leq 2$ and the second term in (\ref{cardyos}) can be dropped for $d>2$
if one is interested in the asymptotic regime.
In fact, the $q\sim 0$ part of the disorder can be eliminated exactly
from (\ref{cardyos}) and leads
only to trivial redefinitions of the correlation functions.
One can perform a translation of the longitudinal
displacement field $u$ by
\begin{equation} \label{translat}
u_\alpha(x) \to u_\alpha(x) + f_\alpha(x)
\end{equation}
where the Fourier transform of $f$ is
\begin{equation}
f_\alpha(q) = \frac{ i \rho_0 q_\alpha V_{q\sim 0}}{c q^2}
\end{equation}
The translation (\ref{translat}) when performed on the replicated form
(\ref{cardyos}) eliminates the long wavelength term but does
leave the cosine term invariant since it is a local transformation.
Note that such a transformation is only possible due to the fact that
the various Fourier components of the disorder are uncorrelated.
The mean squared relative displacement $\tilde{B}(x)$ becomes
\begin{equation}
\tilde{B}(x) = \left. \tilde{B}(x) \right|_{\Delta_0 = 0} +
 \frac{\Delta_0}{m} \int \frac{d^dq}{(2\pi)^d}
\frac{q_\perp^2}{(c q^2)^2} (1-\cos(q x))
\end{equation}
As expected the additional term produces only a subdominant finite correction
above two dimensions. In the following we simply set $\Delta_0 = 0$.

As is obvious from (\ref{leslong}),
higher Fourier components $V_{q \sim K}$ disorder the lattice below $d=4$.
We will now examine the effect of these Fourier components more
quantitatively.

\newpage

\section{Variational Method} \label{simplified}

We now study the Hamiltonian (\ref{cardyos}) using the variational
method introduced by Mezard and Parisi
\cite{mezard_variational_replica}. Hamiltonians with more realistic elastic
energy terms, directly
relevant for experimental systems,
will be considered in sections (\ref{degal3}) and (\ref{degal2}).

\subsection{Derivation of the saddle point equations}

We now look for the best trial Gaussian Hamiltonian $H_0$ in
replica space which approximates (\ref{cardyos}). It has the
general form \cite{mezard_variational_replica}:
\begin{equation} \label{variat}
H_0 = {1 \over 2} \int {d^dq \over (2 \pi)^d} G^{-1}_{ab}(q)
u_a(q) \cdot u_b(-q)
\end{equation}
where the $[G^{-1}]_{ab}(q)$ is a $n$ by $n$ matrix of variational parameters.
Without loss of generality, the matrix $G^{-1}_{ab}(q)$ can be chosen of the
form $G^{-1}_{ab}=  c q^2 \delta_{ab} - \sigma_{ab}$
where the self energy $\sigma_{ab}$ is simply a matrix of constants.
The connected part is defined as $G_c^{-1}(q) = \sum_b G_{ab}^{-1}(q)$.
We obtain by minimization
of the variational free energy $F_{\text{var}}=F_0+\langle H_{eff}-H_0
\rangle_{H_0}$ the saddle point equations:
\begin{equation} \label{bebe}
G_c(q)  =  \frac1{c q^2}, \quad \sigma_{a \ne b} =
 \sum_{K} {\Delta_K \over m T} K^2 e^{- {K^2 \over 2}  B_{ab}(x=0)}
\end{equation}
where $m$ is the number of components of $u$.
One defines
\begin{eqnarray} \label{bebete}
B_{ab}(x) & = & \frac1m \langle [u_a(x) - u_b(0)]^2 \rangle  \\
            & = &  T \int {d^dq \over {(2\pi)}^d}
(G_{aa}(q) + G_{bb}(q) - 2 \cos(qx) G_{ab}(q)) \nonumber
\end{eqnarray}
Note that the connected part is unchanged by disorder,
a direct consequence of the statistical symmetry of (\ref{cardyos})
noted above.
Two general classes of solutions can exist for (\ref{bebe}). One preserves
the symmetry of permutations of replica, and amounts to mimic
the distribution ( thermal and over disorder) of each displacement mode $u(q)$
by a simple Gaussian. The other class, which is a better approximation in the
glassy phase,
breaks replica symmetry and approximates the distribution of displacements
by a hierarchical superposition of gaussians centered at different points in
space
\cite{mezard_variational_replica}. Each gaussian at the lowest level of the
hierarchy corresponds to a different metastable "pinned" position of the
manifold.

\subsection{Replica Symetric solution} \label{repsym}

Let us first examine the replica symmetric solution $G_{a \ne b}(q)=G(q)$ and
$B_{a \ne b}(x)=B(x)$. Using (\ref{bebete}) one has
\begin{equation}
B(x=0) = 2 T \int \frac{d^dq}{(2\pi)^d} G_c(q)
\end{equation}
For $d \le 2$, $B(x=0)$ is infinite, and the off-diagonal part of
$\sigma_{ab}$ is zero.
The $K\ne 0$ Fourier components of the disorder do not contribute. This
solution turns out to be the correct solution for $d < 2$, as
shown in appendix~\ref{stab}. This can be explained physically by the
fact that for $d<2$, thermal fluctuations are strong enough to disorder
the system.

For $d \ge 2$, $\tilde{G}$ is given by
\begin{equation} \label{lala}
\tilde{G}(q) = \frac1{c q^2} + \frac{1}{c^2 q^4 m T} \sum_K \Delta_K
                                    K^2 e^{-K^2 l_T^2/2}
\end{equation}
$l_T$ is the Lindemann length and measure the strength of thermal
fluctuations. It is defined as:
\begin{equation} \label{lala2}
l_T^2 = 2 \langle u^2 \rangle_T = 2 T \int \frac{d^dq}{(2\pi)^d}
\frac1{c q^2} \simeq \frac{2 T S_d}{c (d-2)} (2\pi/a)^{d-2}
\end{equation}
where $1/S_d=2^{d-1}\pi^{d/2}\Gamma[d/2]$.
Due to the term in $1/q^{4}$ in (\ref{lala}), the relative displacement
correlation function grows as
\begin{equation}
\tilde{B}(x) \sim x^{4-d}
\end{equation}
The replica symmetric solution is therefore equivalent at large
distances to the Larkin result based on a model
of independent random forces acting on each vortex. As explained
 in the introduction this
solution does not contain the right physics to describe the long distance
behavior. In this variational approach this shows by the fact that the
replica symmetric solution is unstable towards replica symmetry breaking
for $2<d<4$. This can be checked from the eigenvalue $\lambda$ of
the replicon mode \cite{mezard_variational_replica}.
\begin{equation} \label{replicon}
\lambda=1 - \sum_K
\frac{\Delta_K K^4}{m} e^{- K^2 T \int {d^dp \over (2\pi)^d}
G_c(p) } \int {d^dq \over (2\pi)^d} G_c^2(q)
\end{equation}
A negative
eigenvalue $\lambda$ indicates an instability of the replica symmetric
solution.
We introduce a small regularizing mass in $G_c$: $G_c(q)^{-1}=c q^2 +
\mu^2$ and take the limit $\mu \to 0$. It is easy to see from (\ref{replicon})
that for $d<2$ the replica symmetric solution
is always stable (see also
appendix~\ref{stab}). In that case disorder is in fact irrelevant, due
to the strong thermal fluctuations.
For $d=2$ the condition becomes $\mu^{-2(1-{T K_0^2 \over 4 \pi c})}<1$ for
small $\mu$.
Thus there is a transition
at $T=T_c=4 \pi c/K_0^2 $ between a replica symmetric
stable high temperature phase where disorder is
irrelevant and a low temperature (glassy) phase
where the symmetric saddle point is unstable. We will examine the
physics in $d=2$ in details in section~\ref{degal2}.
For $2<d<4$ the replica symmetric solution is {\it always
unstable} and disorder is therefore always relevant.

\subsection{Replica Symmetry breaking for $2<d<4$ } \label{repsymbreak}

Since for $2 \leq d < 4$, the replica symmetric solution is unstable, to
obtain the correct physics one has to look for a replica symmetry broken
solution. We will focus here on the case $2<d<4$,
the $d=2$ case being discussed in section~\ref{degal2}. Following
\cite{mezard_variational_replica} we denote $\tilde{G}(q)=G_{aa}(q)$,
similarly $\tilde{B}(x) = B_{aa}(x)$, and parametrize $G_{ab}(q)$
by $G(q,v)$ where $0<v<1$, and $B_{ab}(x)$ by $B(x,v)$.
Physically, $v$ parametrises pairs of low lying states,
in the hierarchy of states, as described in \cite{mezard_variational_replica}
$v=0$ corresponding to states further apart.
The saddle point equations become:
\begin{equation} \label{saddle}
\sigma(v)= \sum_K \frac{\Delta_K}{m T} K^2 e^{-{1\over 2} K^2 B(0,v)}
\end{equation}
where
\begin{equation}
B(0,v) = 2T \int \frac{d^dq}{(2\pi)^d} (\tilde{G}(q) -G(q,v))
\end{equation}
$B(0,v)$ corresponds physically to the mean squared relative displacement
between the position of the same vortex ($x=0$) when
the manifold is in two different low lying metastable states.
The large distance behaviour of disorder-averaged correlators
will be determined by the small $v$ behaviour of $B(0,v)$,

As we will show in section~\ref{crosso},
to discuss the large distance behavior $x \gg \xi$ it is enough
to keep the smallest reciprocal lattice vectors with
$K^2=K_0^2$ in (\ref{saddle}) since $B(0,v) \gg a^2$.
We will thus first study a single cosine model obtained by keeping
only $K=K_0$
\begin{equation} \label{singlecosine}
H_{\text{eff}}  =  \frac{c}2 \int d^dx  \sum_{a} (\nabla u^a(x))^2
- \frac{\Delta}{2T} \cos(K_0 (u^{a}(x)-u^{b}(x)))
\end{equation}
Each time we consider this particular model, e.g
in the following subsection (\ref{singlecos}),
we will denote by
$\Delta = N_b \Delta_{K_0}$ where $N_b$ is the coordination number,
i.e the number of vectors $K_0$ with minimal norm.

\subsubsection{Asymptotic behavior (single cosine model)}
\label{singlecos}

We look for a solution such that $\sigma(v)$ is constant for $v>v_c$,
$v_c$ itself beeing a variational parameter, and has an arbitrary
functional form below $v_c$. The algebraic rules for inversion of
hierarchical matrices
\cite{mezard_variational_replica} give:
\begin{equation} \label{inversion}
B(0,v)=B(0,v_c)+ \int_{v}^{v_c} dw \int \frac{d^dq}{(2 \pi)^d}
{2 T \sigma'(w) \over {(G_c(q)^{-1} + [\sigma](w))}^2 }
\end{equation}
where $[\sigma](v)=u\sigma(v)-\int_{0}^{v} dw \sigma(w)$ and
\begin{equation}
B(0,v_c) = \int \frac{d^dq}{(2\pi)^d} \frac{2 T}{G_c(q)^{-1} +
[\sigma](v_c)}
\end{equation}
In that case,
taking the derivative of (\ref{saddle}) ( keeping only $K$=$K_0$ ) with respect
to $v$, using
$[\sigma]'(v)=v\sigma'(v)$, (\ref{inversion}), and
(\ref{saddle}) again
one finds
\begin{equation} \label{relation}
1 = \sigma(v) \int \frac{d^dq}{(2 \pi)^d}
{K_0^2 T \over {(c q^2 + [\sigma](v))}^2 } \simeq
\sigma(v) \left(\frac{T K_0^2 c_d}{c^{d/2}}\right)[\sigma(v)]^{(d-4)/2}
\end{equation}
Since the integral is ultraviolet convergent,
we have taken the short-distance momentum cutoff $\Lambda=2 \pi/a$ to
infinity, a limit discussed below. Derivating one more time
one gets for the effective self energy:
\begin{equation}\label{sigmeq}
[\sigma](v)=  (v/v_0)^{2 / \theta}
\end{equation}
where $v_0 = 2 K_0^2 T c_d c^{-d/2}/(4-d)$ and
\begin{equation}
c_d = \int \frac{d^d q}{(2 \pi)^d} \left(\frac1{q^2+1}\right)^2
= \frac{(2-d) \pi^{1 - d/2}}{2^{d+1} \sin(d\pi/2) \Gamma(d/2)}
\end{equation}
with $c_{d=3}=1/(8 \pi)$, $c_{d=2}=1/(4 \pi)$.

The behavior of $[\sigma](v)$ controls the scaling of the
energy fluctuations \cite{mezard_variational_replica}
$\Delta F \propto L^{\theta} \propto T/v $, with the scale $L$,
and
the large scale behavior is controlled by small $v$.
(\ref{sigmeq}) thus gives an energy fluctuation
exponent $\theta=d-2$.

Using (\ref{sigmeq}) in (\ref{tilde})
one can now compute the correlation functions. Larger distances will
correspond to less massive modes, and one obtains
\begin{eqnarray} & & \label{btildedef}
\overline{ \langle (u(x) - u(0))^2 \rangle } =
2 m T \int {d^dq \over (2\pi)^d}
(1 - \cos(qx) ) \tilde{G}(q) \\
& &
\tilde{G}(q) = { 1\over{c q^2}} ( 1 + \int_{0}^{1} {dv \over v^2}
{ [\sigma](v) \over { c q^2 + [\sigma](v) } } ) \sim \frac{Z_d}{q^d}
\label{gtilde}
\end{eqnarray}
with $Z_d=(4-d)/(TK_0^2 S_d)$ and $1/S_d=2^{d-1}\pi^{d/2}\Gamma[d/2]$.
Thus for $2<d<4$ we find {\it logarithmic} growth
\cite{giamarchi_vortex_short,korshunov_variational_short}:
\begin{equation} \label{meansq}
\overline{ \langle (u(x) - u(0))^2 \rangle }=\frac{2m}{K_0^2} A_d \log|x|
\end{equation}
with $A_d=4-d$. Note that the amplitude is independent of temperature
and disorder.

The solution (\ref{sigmeq}), is a priori valid up to a breakpoint $v_c$,
above which $[\sigma]$ is constant, since $\sigma'(v) = 0$ is also a
solution of the variational equations.
To obtain the behavior at shorter scales for the single cosine
model,
we need to determine the breakpoint $v_c$. For $v>v_c$ using
$[\sigma](v)=\Sigma$ one can rewrite (\ref{sigmeq}) as
\begin{equation}
[\sigma](v)=\Sigma (\frac{v}{v_c})^{\frac{2}{\theta}}
\end{equation}
with $v_c=v_0 \Sigma^{\frac{d-2}{2}}$.
Using (\ref{saddle}) and (\ref{relation})
the equation determining $\Sigma$ is:
\begin{equation} \label{detsigma}
\Sigma^{\frac{(4-d)}{2}}= \frac{\Delta K_0^4 c_d}{m c^{d/2}}
 e^{-\frac12 K_0^2 B(0,v_c)}
\end{equation}
in terms of the nonuniversal quantity $B(0,v_c)$:
\begin{equation} \label{nonuniversal}
B(0,v_c) = 2 T \int \frac{d^dq}{(2\pi)^d} \frac1{c q^2 + \Sigma}
\end{equation}
One can define a length $l$ such that $c l^{-2} = \Sigma$. Since
$x \ll l$ corresponds to $v \gg v_c$ where $[\sigma](v)$ is a constant
and the solution is similar to a
replica symmetric one. $l$ is therefore the length below which the
Larkin regime will be valid. When $l \gg a$, one
finds $B(0,v_c) \approx {l_T}^2$.
For instance in $d=3$, $B(0,v_c)={l_T}^2(1-(a/(2\pi l))\arctan(2
\pi l/a))$. Equation (\ref{detsigma}) can be rewritten
\begin{equation}
1 -  \frac{\Delta K_0^4}{m} e^{- T K_0^2
\int \frac{d^dq}{(2\pi)^d} \frac1{c q^2 + \Sigma}}
      \int \frac{d^dq}{(2\pi)^d} \frac1{(c q^2 + \Sigma)^2} = 0
\end{equation}
which is equivalent to $\lambda_{\rm replicon}(\sigma) = 0$.
Assuming $l \gg a$ one finds
\begin{equation} \label{lvsxi}
l \simeq \xi e^{- K_0^2 {l_T}^2/(4-d)} \simeq \xi
\end{equation}
with
\begin{equation} \label{xiisotrope}
\xi=(m c^2/\Delta K_0^4 c_d)^{1/(4-d)}
\end{equation}
and
\begin{equation} \label{finalvc}
v_c=\frac{2 K_0^2 T c_d}{(4-d) c l^{d-2}}
\end{equation}
Note that although the breakpoint $v_c \to 0$, when $T\to 0$, the length
$l$ which is associated to the transition between the two regimes
remains finite.

For the single cosine model, the characteristic
length $l$ below which the replica symmetric part of the solution
($[\sigma](u)= \Sigma$) determines the physics, is equal to $\xi$
the length for which the relative displacements
are of order $a$.
For this model one has a direct crossover between the Larkin regime and
the logarithmic growth of the displacements.
This is to be expected
since the disorder is here characterized by a single harmonic.
It has therefore no fine structure for distances smaller than $a$, the
lattice spacing. This will not be the case any more if higher harmonics
are included. The disorder will be able to vary strongly for distances
smaller than $a$, and one expects $l$ and $\xi$ to be different, and a
third regime to appear in between: the so called random manifold regime.

\subsubsection{Study of the crossover} \label{crosso}

We study now the full model (\ref{cardyos}). Since this model contains
all the harmonics of the disorder, it can describe correctly the short
distance regimes. In particular, we will examine here the crossover from
the random manifold regime to the logarithmic one.

In order to rewrite the equations in term of
dimensionless quantities, we introduce
the rescaling
\begin{eqnarray} \label{dimensionless}
\sigma(v) & = & \frac{c \xi^{-2}}{v_\xi} s(v/v_\xi) \\
B(v) & = & \frac{a^2}{2 \pi^2} b(v/v_\xi) \nonumber
\end{eqnarray}
as will be obvious later
$\xi$ is the crossover length between the random manifold regime
and the logarithmic one, and $v_\xi$ corresponds to the value of $v$
for which the crossover occurs.
One chooses $\xi$ and $v_\xi$ such that
(\ref{saddle}) and (\ref{inversion})
become in terms of the dimensionless quantities (\ref{dimensionless})
\begin{eqnarray} \label{tobesolved}
s(y) & =  & \sum_p p^2 e^{-p^2 b(y)} \\
\label{threebesolved}
b(y) & = & b(y_c) + \int_y^{y_c} dy
\frac{s'(y)}{[s](y)^{(4-d)/2}}
\end{eqnarray}
where $y_c = v_c/v_\xi$ and the integration over momentum in
(\ref{inversion}) has been performed.
We have introduced the dimensionless variable $p$ such that
$K=2 \pi p/a$.
When using the definition (\ref{tobesolved}) one gets
\begin{eqnarray} \label{lengths}
\xi & =&  \left(\frac{m a^4 c^2}{16 \pi^4 \Delta
                                 c_d}\right)^{\frac1{4-d}}
\\
v_\xi & = &  \frac{2\pi^2}{a^2}\left(\frac{2 T c_d a^{2-d}}{c}\right)
             \left(\frac{a}{\xi}\right)^{d-2}
             \sim \left(\frac{l_T}{a}\right)^2
             \left(\frac{a}{\xi}\right)^{d-2} \nonumber
\end{eqnarray}
Thus $v_\xi$ is always very small compared to $1$. In (\ref{lengths}),
for simplicity,
we have assumed that all $\Delta_K$ have the same value
$\Delta=\Delta_{K_0}$.

The equations (\ref{tobesolved}) and (\ref{threebesolved})
can be solved in a parametric form.
We introduce the variable $z = b(y)$ and define
\begin{equation} \label{deff}
h(z) = \sum_p (p^2)^2 e^{-z p^2} \qquad
H(z) = \sum_p p^2 e^{-z p^2} \qquad \nonumber
\end{equation}
It is possible to keep different $\Delta_K$, for instance
$\Delta_K \sim exp(- K^2 \xi_0^2)$ to describe the effect
of the finite correlation length of the random potential,
by just modifying the functions $H$ and $h=-H'$ to:
\begin{equation}
h(z) = \sum_p \frac{\Delta_K}{\Delta_{K_0}} (p^2)^2 e^{-z p^2}
\end{equation}

Using the variable $z$ and taking the derivative of
(\ref{threebesolved}),
the equations (\ref{tobesolved}) and (\ref{threebesolved}) become
\begin{eqnarray}
s(y) & = & H(z) \label{selfcons1} \\
b'(y) & = & - \frac{s'(y)}{[s](y)^{(4-d)/2}} \label{selfcons2}
\end{eqnarray}
Taking the derivative of (\ref{selfcons1}) one gets
\begin{eqnarray}
s(y) & = & H(z) \\
\ [s](y)^{\frac{4-d}2} & = & h(z)
\end{eqnarray}
Finally, using
$y = \frac{[s]'(y)}{s'(y)} = \frac{d [s]}{dz}/\frac{d s}{d z}$
we obtain the solution in a parametric form
\begin{eqnarray} \label{implicit}
[s](y)^{\frac{4-d}2} & = & h(z) \\
y & = & -\frac2{4-d} h'(z)
        h(z)^{\frac{2d-6}{4-d}} \label{implicit2}
\end{eqnarray}
with $z_c=b(y_c) < z < \infty$.

Let us examine first the various asymptotic behaviors of the solution
(\ref{implicit}). As will be obvious later, large $z$ correspond to
large scales and small $z$ to small scales.
At large $z$  only the smallest $p$ contributes in
the sum (\ref{deff}) for $h(z)$ giving
\begin{eqnarray} \label{largez}
h(z) & = & 2 m e^{-z} \qquad \text{for a square lattice} \\
h(z) & = & \frac{32}3 e^{-4 z/3} \qquad \text{for a triangular
lattice} \nonumber
\end{eqnarray}
In that case the high harmonics are irrelevant and
(\ref{implicit}-\ref{implicit2})
give back formula (\ref{sigmeq}), for the single cosine model.
One recovers the quasi ordered large distance logarithmic regime.

We now study the behavior at small $z$. In that case all harmonics
must be kept and it is convenient to use the following duality
transformation of formula (\ref{deff})
\begin{equation}
I(z) = \sum_{p} e^{-p^2 z} = \frac1{\Omega} \sum_{R}
\left(\frac{\pi}{z}\right)^{m/2} e^{-\frac{\pi^2}{z}R^2}
\end{equation}
where the vector $p$ have been defined above.
$\Omega$ is the volume of the unit cell in the space of the vector $p$
($\Omega = 2/\sqrt{3}$ for a triangular lattice and $1$ for a square
lattice). The vector $R$ are the reciprocal vectors of the $p$ which themselves
are
normalized in units of $2\pi/a$. The $R$ thus correspond to the original
lattice
with a spacing unity.
For small $z$, only $R = 0$ contributes and
\begin{equation}
I(z) \sim \frac1\Omega
\left(\frac{\pi}{z}\right)^{m/2}
\end{equation}
therefore
\begin{equation} \label{smallz}
h(z) \simeq \frac{\pi^{m/2}}{\Omega} \frac{m(m+2)}{4}
\frac1{z^{(m+4)/2}}
\end{equation}
which gives
\begin{equation} \label{randomman}
[s](y) = \Xi y^{2/\theta_{\text{RM}}}
\end{equation}
where the fluctuation energy exponent of the
random manifold regime is $\theta = (2d -2m + d m)/(4+m)$ and the
amplitude
\begin{equation}
\Xi =
\left(\frac{\pi^{m/2}}{\Omega}\frac{m(m+2)}{4}\right)^{\frac{4}{2d-2m+dm}}
\left(\frac{4-d}{4+m}\right)^{2/\theta_{\text{RM}}}
\end{equation}

The equations (\ref{tilde}) and (\ref{gtilde})
once rescaled using (\ref{lengths}) give
\begin{equation} \label{rescaledbtilde}
\tilde{B}(x) = \frac{a^2}{2\pi^2} \tilde{b}( \frac{x}{\xi})
\end{equation}
and
\begin{equation}
\tilde{b}(x) = \frac{\pi
\theta_{\text{RM}}}{2\sin(\pi\theta_{\text{RM}}/2) c_d}
\Xi^{\theta_{\text{RM}}/2}
\int \frac{d^dq}{(2\pi)^d}
\frac1{q^{2+\theta_{\text{RM}}}} (1-\cos(qx))
\end{equation}
with the useful intermediate formula:
\begin{equation}
\int_0^{\infty} \frac{dy}{y^2} \frac{y^a}{1+y^a} = \frac{ \pi/a }{\sin{\pi/a}}
\end{equation}

Using the asymptotic expression
\begin{equation}
\int \frac{d^dq}{(2\pi)^d} \frac{1}{q^{d+2\nu}} (1-\cos(q x))
     = I_{d,\nu} x^{2\nu}
\end{equation}
with
\begin{equation}
I_{d,\nu} = \frac{\pi^{1-d/2}}{2^{d+2\nu}
                       \Gamma(d/2+\nu)\Gamma(1+\nu)\sin(\nu\pi)}
\end{equation}
one gets
\begin{equation} \label{finalrandom}
\tilde{b}(x) \sim \left(\frac1{c_d}
\frac{\pi\theta_{\text{RM}}}{2\sin(\pi\theta_{\text{RM}}/2)}
\Xi^{\theta_{\text{RM}}/2} I_{d,\nu} \right) x^{2\nu}
\end{equation}
with $2\nu=2+\theta_{\text{RM}}-d$.
Thus the exponent entering in the relative diplacement growths is
$\nu = \frac{4-d}{4 + m}$ with $\nu=1/6$ for $m=2$.
This corresponds to the random manifold regime
\cite{bouchaud_variational_vortex_prl,bouchaud_variational_vortex}.
In this regime each vortex is held by the elastic forces of the other
vortices and sees an independent random potential. The mean squared
displacement
grows more slowly than in the Larkin regime (the exponent is $1/3$,
compared to $1$ for the Larkin regime in $d=3$).
In $d=3$ and $m=2$ one gets
$\theta_{\text{RM}}=4/3$ and for a triangular lattice in $d=3$
\begin{equation}
\Xi = \frac16 \sqrt{\frac{\pi}{3\Omega}}
\end{equation}
the amplitude in (\ref{finalrandom}) is $\simeq 2.3817$.

As for the single cosine model of section~\ref{singlecos}, the solution
(\ref{randomman}) is valid up to a breakpoint $v_c$ above which the self
energy $[\sigma](v)$ is constant. This corresponds to
scales such that $\tilde{B}(x)$ is smaller than $l_T^2$ and $\xi_0^2$.
One then recovers the replica symmetric propagator $\tilde{G}(q)
\sim 1/q^4$ for $q^2 \gg [\sigma](v_c)$, and Larkin's model behavior.
To compute the crossover
and to determine the breakpoint $v_c=y_c v_\xi$, we
proceed similarly to section~\ref{singlecos}.
The equation determining $\Sigma$ is now
\begin{equation} \label{breakpoint}
\Sigma^{\frac{(4-d)}{2}}= \sum_{K} \frac{\Delta_K K^4 c_d}{m c^{d/2}}
 e^{-\frac12 K^2 B(0,v_c)}
\end{equation}
where $B(0,v_c)$ is given in (\ref{nonuniversal}).
Note that keeping the correlation length of the disorder
using $\Delta_K = \Delta \exp(-1/2 K^2 {\xi_0}^2)$ amounts
to change $B(0,v_c)$ into $B(0,v_c) + {\xi_0}^2$ and thus
$l_T^2$ into $l_T^2 + {\xi_0}^2$. We will thus take
$\Delta_K = \Delta$ keeping in mind this change.

Solving (\ref{breakpoint}) and using the small $z$ expansion of $h(z)$
one gets for the lengthscale  $l$ such that $\Sigma = c l^{-2}$(assuming
$l \gg a$ and $l_T \ll a$)
\begin{equation} \label{complique}
l = C_s \xi \left(\frac{l_T}{a} \right)^{\frac{1}{\nu}}
\end{equation}
where
\begin{equation}
C_s = (2 \pi^2)^{1/(2\nu)}(\frac{4\Omega}{\pi^{m/2}m(m+2)})^{1/(4-d)}
\end{equation}
The breakpoint $v_c$ can obtained using (\ref{implicit}) with the
value $z=z_c = 2\pi^2 l_T^2/a^2$. This gives using (\ref{lengths})
\begin{equation}
y_c \sim \left(\frac{l_T}{a}\right)^{-\frac{\theta_{\text{RM}}}{\nu}}
\end{equation}
which leads to
\begin{equation} \label{nobreakpoint}
v_c \sim \left(\frac{a/\xi}{(l_T/a)^{1/\nu}}\right)^{d-2}
    \sim \left(\frac{a}{l}\right)^{d-2}
\end{equation}
The characteristic length $l$ separates the Larkin regime from the
random manifold regime, and is in that case much smaller than $\xi$.
Lowering the temperature reduces the range over which the Larkin
regime occurs. This is because the thermal wandering responsible for
smoothing the random potential on a scale $l_T$ decreases.
The relative displacements of two vortices separated by
$l$ is of order
$\text{Max}(l_T,\xi_0)$,
giving $\tilde{B}(l) \sim \text{Max}(l_T,\xi_0)^2$.
As $T\to 0$, $l$ becomes identical to the Larkin-Ovchinikov length $R_c$.
Using the expression of
$\tilde{B}(x) \sim C x^{2\nu}$
in the random manifold regime and the additional relation $\tilde{B}(\xi)
\sim a^2$ one recovers the expression (\ref{complique}) for $l$.
When $v_c = 1$ the Larkin regime disappears.
This occurs when $l \sim a$.
The criterion $l \gg a$ for which the Larkin regime exists
is equivalent to
\begin{equation} \label{condition}
\frac{a}{\xi} \ll \left(\frac{l_T}{a}\right)^{\frac1{\nu}}
\end{equation}
and corresponds therefore to extremely weak disorder and intermediate
temperatures. The absence of a Larkin regime means that the disorder-induced
relative displacement of two neighbors in the lattice is already larger
than $\text{Max}(l_T,\xi_0)$.

\subsubsection{Crossover in $d=3$}

In $d=3$, it is possible to solve the equations describing the crossover
analytically, and thus to obtain the full crossover function between the
random manifold and the quasi ordered regime. We will
examine $d=3$ and $m=2$.
for the model (\ref{cardyos}). Such a case is physically relevant for
the case of vortex lattices. The crossover length are given by
(\ref{lengths})
\begin{eqnarray}
\xi & = & \frac{a^4 c^2}{\pi^3 \Delta} \\
v_\xi & = & \frac{\pi^4}2 T \Delta/(a^6 c^3) \sim l_T^2/(a \xi)
\nonumber
\end{eqnarray}
Using (\ref{rescaledbtilde}) one has
\begin{equation} \label{audessus}
\tilde{b}(x) = \frac1{c_3} \int \frac{d^3 k}{(2\pi)^3} (1-\cos(k x))
\int_0^{y_c} \frac{dy}{y^2} \frac{[s](y)}{k^2(k^2 + [s](y))}
\end{equation}
Performing the angular integration over momentum in (\ref{audessus})
we find
\begin{equation}
\tilde{b}(x) = \frac{1}{2\pi^2 c_3} \int_0^\infty dk  (1-\frac1{k |x|}
\sin(k |x|))
\int_0^{\infty} \frac{dy}{y^2} \frac{[s](y)}{k^2 + [s](y)}
\end{equation}
where we have extended the integral over $y$ to infinity,
assuming $v_\xi \ll v_c$ or equivalently $l_T \ll a$, in
which case there is a wide random manifold regime.
Performing the remaining integration over $k$ one gets
\begin{equation}
\tilde{b}(x) =
\frac{1}{4\pi c_3} \int_0^\infty \frac{dy}{y^2} \left( [s]^{1/2}(y) -
\frac1{|x|}(1-e^{-|x| [s]^{1/2}(y)})\right)
\end{equation}
Using the parametric solution (\ref{implicit}) and (\ref{implicit2})
for $[s](y)$ we obtain the final expression
\begin{eqnarray} \label{bani}
\tilde{b}(x)
& = & \int_0^\infty dt \frac{h''(t) h(t)}{h'(t)^2} f(x h(t))   \\
& & f(x) = 1 - \frac1x (1 - e^{-x})
\end{eqnarray}
Expression (\ref{bani}) gives the full relative displacements
correlation function as a function of distance. To recover the
asymptotic expression of section~\ref{singlecos}, for large distance
$x$, one notices that
in (\ref{bani}), $h(t)\sim A e^{-\alpha t}$ as shown
in (\ref{largez}) for large $t$. Thus the large $x$ behavior will be
controlled by small $t$. One obtains the asymptotic expression
\begin{equation}  \label{asympisoan}
\tilde{b}(x) = \frac1\alpha\left[f(\infty) (\log(A x)+1)
               + \int_0^{\infty} dz \log(1/z) f'(z) \right]
\end{equation}
where we have used $[h/h']_0^{\infty}=-1/\alpha$.
Using (\ref{bani}), one finds
\begin{equation} \label{asympisofin}
\tilde{b}(x) = \frac1\alpha\left[\log(A x) + \gamma \right]
\end{equation}
where $\alpha = [K_0 a /(2\pi)]^2$ and $\gamma \simeq 0.57721$ is the Euler
constant.
This implies that the translational order correlation function
\begin{equation}
C_{K_0}(x) = \langle e^{i K_0 \cdot u(x)} e^{-i K_0 \cdot u(0)} \rangle
\end{equation}
behaves for large $x$ as:
\begin{equation}
C_{K_0}(x) = \frac{e^{-\gamma}\xi}{A |x|}
\end{equation}
One recovers the power law behavior of section~\ref{singlecos} as well
as the amplitude. The intermediate distance behavior will be examined in
more details for more realistic elastic Hamiltonians in connection with
vortex lattices, in section~\ref{degal3}.

\subsection{Self-consistence of the physical asumptions}
\label{selfconsist}

Finally,
for our solution to be valid, one has to check self-consistently that
even in the presence of disorder, the basic assumption that elastic
theory was applicable remains valid. One has therefore to check that
$\frac1m \overline{\langle (\nabla u)^2 \rangle} \ll 1$.

For simplicity we will make the analysis for the single cosine model
(for which $l\sim \xi$) but similar results can be derived for the full
Hamiltonian. Using the variational solution of section~\ref{singlecos},
one obtains
\begin{equation} \label{departself}
\frac1m \overline{\langle (\nabla u)^2 \rangle}
= \frac{2 T}{c}
\int \frac{d^d q}{(2 \pi)^d} \left[\int_0^{v_c} \frac{dv}{v^2}
\frac{\Sigma (v/v_c)^{2/\theta}}{c q^2 + \Sigma(v/v_c)^{2/\theta}} +
(\frac{1}{v_c}-1)\frac{\Sigma}{c q^2 + \Sigma}  + 1 \right]
\end{equation}
Using (\ref{lala2}) one gets, for the case $l \gg a$
\begin{equation}
\frac1m \overline{\langle (\nabla u)^2 \rangle}
\simeq \Sigma \frac{l_T^2}{c v_c}\frac{2}{4-d} +
(\frac{2\pi l_T}{a})^2 \frac{d-2}d
\end{equation}
where the last contribution is due to thermal fluctuations only.
Replacing $v_c$ by  (\ref{finalvc}) one finds
\begin{eqnarray} \label{arrivee}
\overline{\langle (\nabla u)^2 \rangle}
& = & \frac{8\sin(\pi(d-2)/2)}{\pi(d-2)^2} \left(\frac{a}{2\pi
                                             l}\right)^{4-d} +
                            (\frac{2\pi l_T}{a})^2 \frac{d-2}d
\nonumber
\end{eqnarray}
When $d\to 2$ one cannot neglect $\Sigma$ in the denominator of
(\ref{departself}), and the expression (\ref{arrivee}) becomes
\begin{equation}
\overline{\langle (\nabla u)^2 \rangle} \sim \left(\frac{a}{l}\right)^2
\log(l/a)
\end{equation}
As is obvious from (\ref{arrivee}), one has always
$\overline{\langle (\nabla u)^2 \rangle} \ll 1$, provided that $l \gg a$
or equivalently using (\ref{lvsxi}) for the single cosine model,
provided that one is far from the melting temperature
$l_T \ll a$ and that the disorder is weak $\xi \gg a$. In that case one
can indeed use an elastic theory in the absence of dislocation, even in
the presence of disorder, and our solution is valid in such a
regime.

\subsection{Comparison with BMY}

The previous application of the variational method by BMY
\cite{bouchaud_variational_vortex_prl,bouchaud_variational_vortex}
led to
the erroneous conclusion that the fluctuations are enhanced at
large scale. They find for $d=3$
$B(x) \sim x^{1/2}$ instead of logarithmic behavior found here.
Although they want to describe the same physical situation as the one
studied here,
they in fact consider a model which turns out to be
fundamentally different, in which each vortex sees a different disorder.
In their model the random potential is also dependent on the line
index $i$ such that
\begin{equation} \label{unphysical}
\overline{V(R_i,r,z)V(R_j,r',z')} =\Delta
\delta(r-r')\delta(z-z')|R_i-R_j|^{-\lambda}
\end{equation}
This amounts to introduce an extra
disorder in the original model (\ref{total}) with correlations decaying
as $1/|R_i-R_j|^\lambda$. Then BMY retain only the long wavelength
part of this disorder, which indeed for a fixed
$\lambda>0$ dominates the contribution of higher harmonics. They
then look at the limit of the exponents when $\lambda \to 0$.
The result they obtain with this procedure
is incorrect (although their derivation is
technically sound) and comes from the following artifact: by assuming
that different lines see different random potentials, they make it
possible to optimize the pinning energy by a global translation of the
whole lattice. In that case the pinning energy will obviously be
dependent on $u$, even for a uniform $u$. On the other hand for the
genuine disorder, which is only dependent on the space position of the
lines, it is obvious that a translation by one of the vectors of the
lattice cannot change the energy, and therefore the $q\sim 0 $ part of
$H_{\text{pin}}$ cannot depend on $u$ but only in $\partial u$. By
regularizing the integrals with a disorder dependent on the line index
they introduce an extra and non-physical disorder which is relevant and
changes the long-range behavior of the correlation function compared to
the physical case. Indeed there is a
crossover length $\xi_\lambda$ associated with this disorder above
which the long distance behavior is the one given by BMY. Below this
distance the vortices all see the same random potential. To recover
the physical model one has to take
$\lambda\to 0$ and in that case $\xi_\lambda \to \infty$.

In more mathematical terms, the variational method gives three types
of contributions for the self energy as shown in
appendix~\ref{bmy}
\begin{equation}
\sigma(q,v) \sim c_1 q^2 + c_2 e^{-\frac{K_0^2}2 B(0,v)} +
\lambda B(0,v)^{-1-\lambda/2}
\end{equation}
The first term is the long wavelength contribution of the genuine
disorder which is irrelevant. The second one is the higher harmonic
contribution which is responsible for the logarithmic growth at large
distances, and
the third term is the long wavelength contribution of this extra
disorder giving a $B(x) \sim x^{1/2}$ for $x > \xi_\lambda$. Only the
third term was kept by BMY, artificially taking the limit $\lambda \to
0$ in the exponent only but {\bf not} in the amplitude of such a term.
Note that if one takes the limit $\lambda\to 0$
(which corresponds to the physical situation) before taking the limit $x\to
\infty$ one recovers that the $q\simeq 0$ part of the disorder does not
play any role, and the amplitude they obtain vanishes.

A simple Flory argument can be made to estimate the effect of the long
wavelength part of the disorder on the displacements.
This confirm that it is irrelevant above
$d=2$ (see also section~\ref{dimensionarg}).
Let $u$ be the typical relative
displacement over a lengthscale $L$. The elastic energy cost is
$u^2 L^{d-2}$ while the typical energy gain due to the disorder is
\begin{equation}
\sqrt{L^d + u L^{d-1}} - \sqrt{L^d} \propto u L^{d/2 -1}
\end{equation}
which comes from the change of density of the vortices. Since the
vortex in the center are unaffected the gain of energy can come only
from {\it boundary} terms. Balancing the two terms one finds
$u \sim L^{(d-2)/2}$
which is obviously irrelevant above two dimensions.

In fact one can simplify the
the saddle point equations of
\cite{bouchaud_variational_vortex_prl,bouchaud_variational_vortex}
by noting that the $x$ dependence of $B(x,u)$ in these equations
is unimportant, up to higher order terms in $\nabla u$.
Such a calculation is performed in the appendix~\ref{bmy}. One then
recovers the local model
(\ref{cardyos}) which is simple and physically transparent
enough to allow for the exact solution of section~\ref{repsymbreak}.

\newpage

\section{Flux Lattices} \label{degal3}

\subsection{Model}

The theory developed in section~\ref{simplified}, when specialized to
$m=2$ and $d=3$, can be applied to
describe the effects of weak disorder on the Abrikosov phase of type II
superconductors.
High T$_c$ superconductors can be modeled by stacks of coupled planes.
The system is therefore described by layers of two dimensional
triangular lattices of vortices.
The displacements $u$ are two dimensional
vectors, hence $m=2$ (the vortex can only move within the plane). We
denote
by $R_i$ the equilibrium position of the vortex labeled by an integer $i$,
in the $x-y$ plane, and by $u(R_i,z)$ their in-plane displacements.
$z$ denotes the coordinate perpendicular to the planes and along the
magnetic field.
The total energy is:
\widetext
\begin{equation} \label{vordep}
H = \frac12 \int d^2r dz [(c_{11}-c_{66})(\partial_\alpha u_\alpha)^2
+c_{66} (\partial_\alpha u_\beta)^2 + c_{44}
(\partial_z u_\alpha)^2]  + \int d^2r dz V(r,z) \rho(r,z)
\end{equation}
\narrowtext
where $\alpha,\beta$ denote in-plane coordinates. The Hamiltonian
(\ref{vordep}) is identical to (\ref{real}) with
$ \Phi_{\alpha\beta}(q) =
G^{-1,T}_c(q) P^T_{\alpha\beta} + G^{-1,L}_c(q) P^L_{\alpha\beta}$
where
\begin{eqnarray} \label{lesgmoinsun}
G^{-1,T}_c(q) & = & c_{44} q_z^2 + c_{66} q_{\perp}^2 \\
G^{-1,L}_c(q) & = & c_{44} q_z^2 + c_{11} q_{\perp}^2  \nonumber
\end{eqnarray}
and
\begin{eqnarray} \label{projectors}
P^T_{\alpha\beta}(q) & = & \delta_{\alpha\beta} - q_\alpha q_\beta /
q_{\perp}^2 \\
P^L_{\alpha\beta}(q) & = & q_\alpha q_\beta / q_{\perp}^2 \nonumber
\end{eqnarray}
are the transverse
and longitudinal propagator. $q_{\perp}$ denotes the in plane vector, whereas
$q_z$ is the out of plane component.
Equation (\ref{vordep}) corresponds
to a local elastic theory, but non local elasticity can also be considered
at the expense of introducing $q$ dependent coefficients $c$. This point
will be considered in greater detail below.
For the moment we restrict ourselves to dispersionless elastic
constants.

Weak point-like disorder such as oxygen vacancies, or defects
introduced artificially in a controlled way, e.g by electron irradiation
\cite{kwok_electron_defects}, can be modeled by a gaussian
random potential of correlation length of order $\xi_0$. Here
the disorder will be taken as completely uncorrelated from plane to
plane $\Delta(x-x')=\Delta(r-r')\delta(z-z')$.
Such a description will be valid as long as
each pinning center is weak enough so that the pinning length $l$
(also called $R_c$ in formula (51) of the Larkin-Ovchinikov paper
\cite{larkin_ovchinnikov_pinning})
is much larger than the average distance between impurities.

The disorder term in (\ref{vordep}) is transformed into a form similar
to (\ref{cardyos}). We can now use the methods of
section~\ref{simplified} with the realistic elastic Hamiltonian
(\ref{vordep}) to get the physical properties of a vortex lattice.
Most of the theoretical calculation are confined in
section~\ref{theory}, whereas a simple physical interpretation of the
results is given in section~\ref{discussion}.
The experimental consequences are
discussed in details in section~\ref{experiments}.

\subsection{Theoretical predictions} \label{theory}

One can then perform a variational ansatz identical
to (\ref{variat}) but for the introduction of the longitudinal
$G^L_{\alpha\beta}$ and transverse part $G^T_{\alpha\beta}$.
\begin{equation}
H_0 = {1 \over 2} \int {d^dq \over (2 \pi)^d} G^{-1}_{\alpha\beta,ab}(q)
u_{\alpha}^a(q) u_{\beta}^b(-q)
\end{equation}
and
\begin{equation}
G_{\alpha\beta,ab}(q) =
G^T_{ab}(q) P^T_{\alpha\beta} + G^L_{ab}(q) P^L_{\alpha\beta}
\end{equation}
The saddle point equation (\ref{saddle}) for the self energy now becomes
\begin{equation} \label{saddleflux}
\sigma_{\alpha\beta}(v)= \frac{\Delta}{m T} \sum_K  K_\alpha K_\beta
e^{-{1\over 2} K_\gamma K_\delta  B_{\gamma\delta}(x=0,v)}
\end{equation}
The correlation function $B_{\alpha\beta}$ is defined by
(\ref{bebete}) with the replacement of $G$ by $G_{\alpha\beta}$.

Since $B_{\gamma\delta}(x=0,v)$ is a purely local quantity it is
isotropic and $B_{\gamma\delta}(x=0,v) =
\delta_{\gamma\delta}B(x=0,v)$. This implies that
$\sigma_{\alpha\beta} = \sigma(v) \delta_{\alpha\beta}$, i.e. an
isotropic self energy. Thus $B(v) = \frac12[B^L(v) + B^T(v)]$, where
by definition
$B^{L,T}(v)$ satisfy equation (\ref{inversion}) with respect to
$G_c^{L,T}$.
Integration over $q$ leads to
\begin{equation} \label{saddleflux2}
B(v) = B(v_c) + T c_d
\frac1{c_{44}^{1/2}}(\frac1{c_{66}}+\frac1{c_{11}})
\int_u^{u_c} \frac{\sigma'(v) dv}{[\sigma]^{1/2}(v)}
\end{equation}
The solution of (\ref{saddleflux}) can trivially be obtained from the
isotropic solution  with the replacement
\begin{eqnarray} \label{replacement}
c & \to & c_{44} \\
c_d & \to & c'_d = c_d
\frac12(\frac{c_{44}}{c_{66}}+\frac{c_{44}}{c_{11}})
\nonumber
\end{eqnarray}
One can now compute the correlation functions
\begin{eqnarray} \label{lesb}
\tilde{B}_{\alpha\beta}(x)
         & = & \overline{\langle [u_\alpha(x) - u_\alpha(0)]
                        [u_\beta(x) - u_\beta(0)] \rangle } \\
         & = &  \tilde{B}^T P^T_{\alpha\beta}(r)+
                \tilde{B}^L P^L_{\alpha\beta}(r) \nonumber
\end{eqnarray}
where the longitudinal and transverse propagators have been defined in
(\ref{projectors}) and similarly to (\ref{btildedef})
\begin{eqnarray} \label{btrans}
\tilde{B}^L(r,z) &=& 2T \int \frac{d^3q}{(2\pi)^3} (\hat{q_{\perp}}\hat{r})^2
                       (1-\cos(q_{\perp} r + q_z z)) \tilde{G}^L(q) + \\
 & &                   [1-(\hat{q_{\perp}}\hat{r})^2]
                       (1-\cos(q_{\perp} r + q_z z)) \tilde{G}^T(q)
\nonumber
\end{eqnarray}
and a similar equation for $\tilde{B}^T$ obtained from (\ref{btrans})
by permuting $L,T$.
$\tilde{G}^{L,T}$ are defined similarly to equation (\ref{gtilde})
with the replacement of $c q^2$ by $G^{-1,LT}$ defined in
(\ref{lesgmoinsun}).
One then rescales $q$ and $r$ to obtain isotropic integrals over
momenta. Equation (\ref{btrans}) then takes the form
\begin{equation}
\tilde{B}^L(r,z) =
\frac{a^2}{2\pi^2}[\frac{c_{44}}{c_{11}}
\tilde{F}^{L}(r\sqrt{\frac{c_{44}}{c_{11}}},z) +
\frac{c_{44}}{c_{66}}
\tilde{F}^{T}(r\sqrt{\frac{c_{44}}{c_{66}}},z)]
\end{equation}
Those integrals contain the self energy $[\sigma](v)$ which is
determined itself from equation (\ref{saddleflux}-\ref{saddleflux2}).
These equations are rescaled similarly to (\ref{dimensionless}) with the
replacement (\ref{replacement}). This defines two crossover length
$\xi$ and $\xi_z$ for in plane and $z$ directions, given by
\begin{equation} \label{crosslength}
\xi_z = \frac{2 a^4 c_{44}}{\pi^3
\Delta(\frac1{c_{66}} + \frac1{c_{11}})}, \qquad
\xi = \sqrt{\frac{c_{66}}{c_{44}}} \xi_z =
\frac{2 a^4 c_{44}^{1/2} c_{66}^{3/2}}{\pi^3 \Delta (1 +
\frac{c_{66}}{c_{11}})}
\end{equation}
where, as we recall $\Delta$ is the disorder strength $\rho_0
\overline{V(q)V(-q)}=\Delta$.

Rescaling by the length $\xi$ and $\xi_z$ one gets
\begin{equation}
\tilde{B}^L(r,z) =
\frac{a^2}{2\pi^2}[\frac{c_{44}}{c_{11}}
\tilde{F}^{L}(\frac{r}\xi \sqrt{\frac{c_{66}}{c_{11}}},\frac{z}{\xi_z})
+ \frac{c_{44}}{c_{66}} \tilde{F}^{T}(\frac{r}{\xi},\frac{z}{\xi_z})]
\end{equation}
where the functions $F$ are given, for $z=0$ by
\begin{eqnarray}
\tilde{F}^{L}(r) & = & \frac1{(2\pi)^3 c'_3} \int_0^1 \frac{dy}{y^2}
        [s](y) \int d^2q_{\perp} dq_z
         \cos^2(\theta)(1-\cos(q_{\perp} r \cos(\theta)))
\frac1{q_{\perp}^2+q_z^2}
         \frac1{q_{\perp}^2+q_z^2+[s](y)} \\
\tilde{F}^{T}(r) & = & \frac1{(2\pi)^3 c'_3} \int_0^1 \frac{dy}{y^2}
        [s](y) \int d^2q_{\perp} dq_z
         (1-\cos^2(\theta))
         (1-\cos(q_{\perp} r \cos(\theta))) \frac1{q_{\perp}^2+q_z^2}
         \frac1{q_{\perp}^2+q_z^2+[s](y)}
\end{eqnarray}
performing the $q_{\perp},q_z$ integrations, one gets
\begin{equation}
\tilde{F}^{T,L} = \frac1{4\pi c'_3}\int_0^\infty \frac{dy}{y^2}
f^{T,L}(y)\\
\end{equation}
where
\begin{eqnarray}
f^L(x) & = & \left[\frac{[s]^{1/2}}2 -\frac1{[s]^{1/2}|x|^2}
             +\left(\frac1{|x|} +
             \frac1{[s]^{1/2}|x|^2}\right)
             e^{-[s]^{1/2}|x|}\right]
             \nonumber \\
f_T(x) & = & f_I(x) - f_L(x) \\
f_I(x) & = & \left[[s]^{1/2} - \frac1{|x|} +
             \frac1{|x|}e^{-[s]^{1/2}|x|}\right] \nonumber
\end{eqnarray}
Expressing again the $[s]$ in term of the functions $h$, one gets
\begin{equation} \label{resdebut}
\tilde{B}^{L,T}(r) = \frac{a^2}{2\pi^2}[
\frac{2 c_{11}}{c_{11}+c_{66}}
\tilde{b}^{T,L}(\frac{r}{\xi}) +
\frac{2 c_{66}}{c_{11}+c_{66}}
\tilde{b}^{L,T}(\sqrt{\frac{c_{66}}{c_{11}}}\frac{r}{\xi})]
\end{equation}
where $\tilde{b}^{L,T}$ have an expression similar
to (\ref{bani})
\begin{equation} \label{simibani}
\tilde{b}^{L,T}(x) =
\int_0^\infty dt \frac{h''(t) h(t)}{h'(t)^2} f^{L,T}(x h(t))
\end{equation}
with $f^{L,T}$ given by
\begin{eqnarray} \label{resfin}
f^{L}(x) & = & \frac12 - \frac1{x^2} + (\frac1x + \frac1{x^2}) e^{-x} \\
f^{T}(x) & = & f(x) - f^L(x)  \nonumber
\end{eqnarray}
(\ref{resdebut}-\ref{resfin}) give the complete expression
of the displacement correlation
function for equal $z$ as a function of the distance
in the transverse plane.

In the large distance regime one obtains an expression similar to
(\ref{asympisoan}), with the $f$ replaced by $f^{T,L}$. This gives
\begin{eqnarray}
\tilde{b}^{L}(x) & = & \frac1{2\alpha} \left[ \log(A x) + \gamma
                       + \frac12 \right] \\
\tilde{b}^{T}(x) & = & \frac1{2\alpha} \left[ \log(A x) + \gamma
                       - \frac12 \right] \nonumber
\end{eqnarray}
where $\alpha = [K_0 a /(2\pi)]^2$, $\gamma \simeq 0.57721$ is the Euler
constant,
and $A=32/3$ for the triangular lattice.
This leads for the $\tilde{B}^{L,T}$ functions at large distance
\begin{equation}
\tilde{B}^{L,T}(r)  =  \frac{2}{K_0^2}
                       \left[ \log(A r/\xi) + \gamma
    + \frac{\epsilon^{L,T}}2 \frac{c_{11}-c_{66}}{c_{11}+c_{66}}
                       + \frac{c_{66}}{2(c_{11}+c_{66})}
                          \log(\frac{c_{66}}{c_{11}}) \right]
\end{equation}
where $\epsilon^L=-1$, $\epsilon^T = +1$. Note that
$\tilde{B}=(\tilde{B}^{L}+\tilde{B}^{T})/2$.

It is interesting to note that complete isotropy, in the
displacement correlation functions, is recovered at large scales.
The translational correlation function is
\begin{equation} \label{discor}
C_{K_0}(r) = \frac{e^{-\gamma}\xi}{A r}
e^{-\frac{c_{66}\log(c_{66}/c_{11})}{2(c_{11}+c_{66})}}
e^{((\hat{K_0}\cdot\hat{r})^2-\frac12)
\left(\frac{c_{11}-c_{66}}{c_{11}+c_{66}}\right)}
\end{equation}

\begin{figure}[tbh]
\plotfig{l}{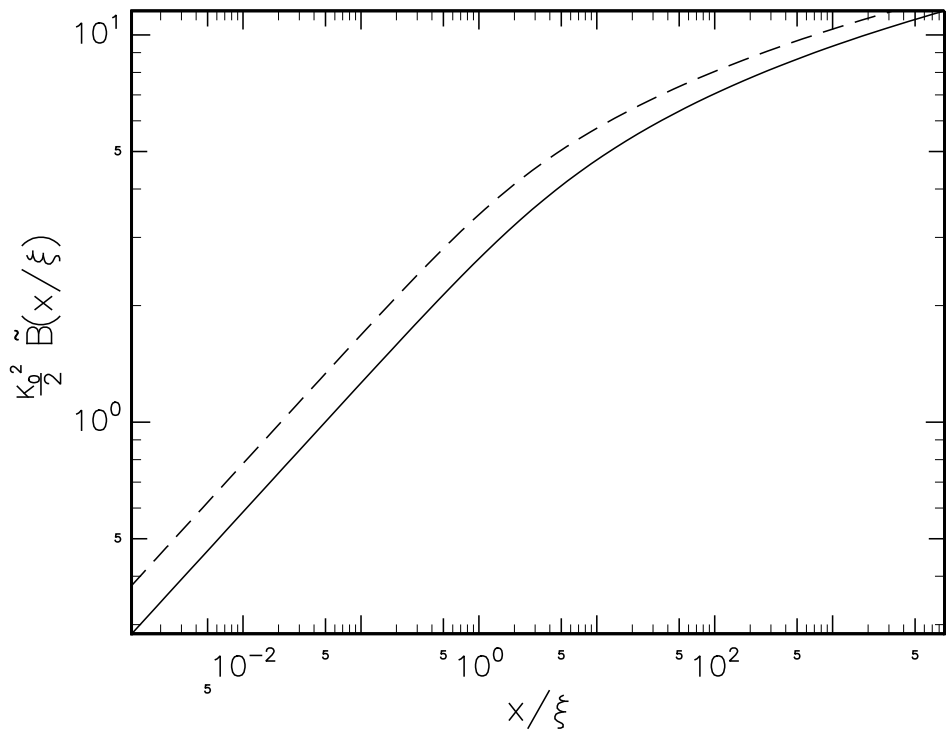}
\caption{\label{figure1}
Plot of $\frac{K_0 ^2}2\tilde{B}_{L,T}$ for the triangular Abrikosov
lattice, in the limit $c_{66} \ll c_{11}$.
The longitudinal (solid line) and
transverse (dashed line) relative displacement correlation functions
$\tilde{B}_{L,T}$ are defined in (\protect{\ref{lesb}}). $K_0$ is one of
the first reciprocal lattice vectors, and $\xi$ is the crossover length
defined in (\protect{\ref{crosslength}}). When $x < \xi$ one sees a
power law with $2\nu=1/3$ characteristic of the random manifold regime.}
\end{figure}

\begin{figure}[tbh]
\plotfig{l}{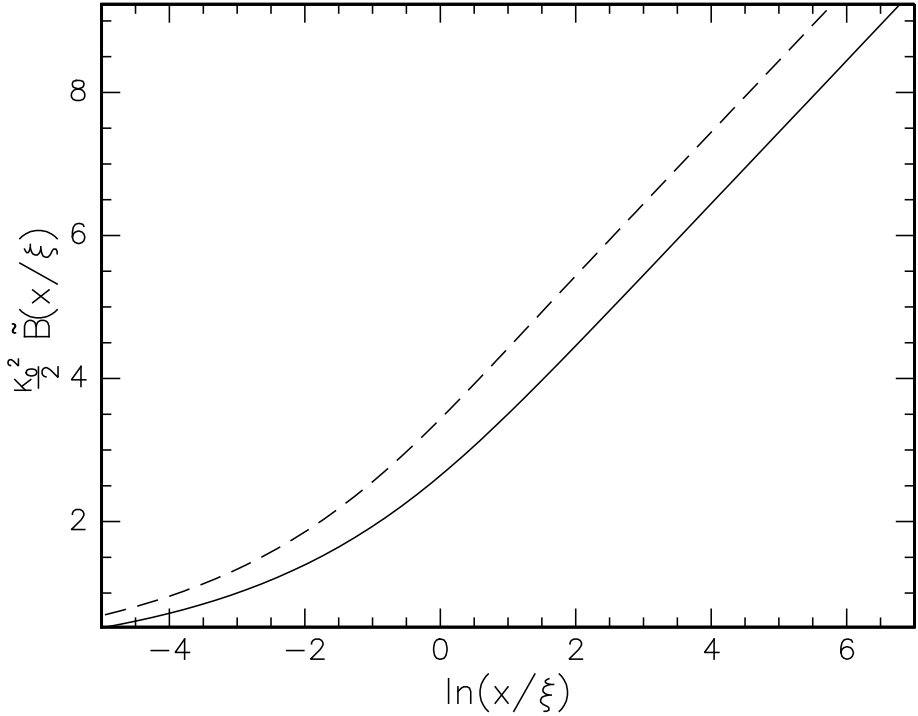}
\caption{\label{figure2}
Plot of $\frac{K_0 ^2}2\tilde{B}_{L,T}$ for the triangular Abrikosov
lattice, in the limit $c_{66} \ll c_{11}$.
The longitudinal (solid line) and
transverse (dashed line) relative displacement correlation functions
$\tilde{B}_{L,T}$ are defined in (\protect{\ref{lesb}}). $K_0$ is one of
the first reciprocal lattice vectors, and $\xi$ is the crossover length
defined in (\protect{\ref{crosslength}}). When $x > \xi$ one sees
the logarithmic regime.}
\end{figure}

\begin{figure}[tbh]
\plotfig{l}{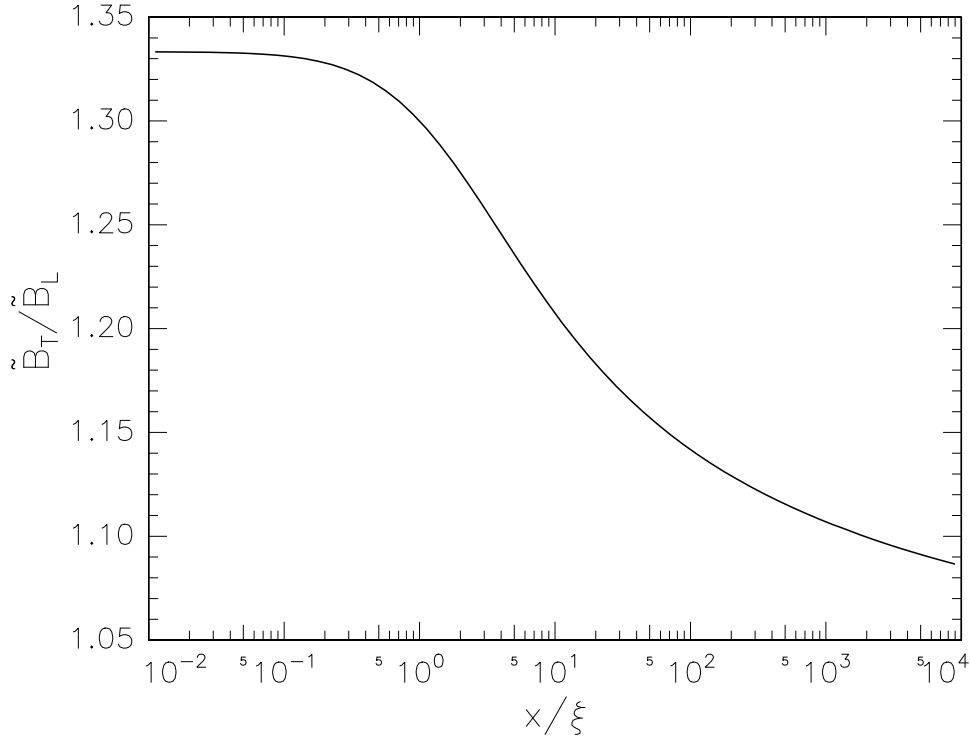}
\caption{\label{figure3}
Plot of the ratio
$\tilde{B}_{T}/\tilde{B}_{L}$ for the triangular Abrikosov
lattice, in the limit $c_{66} \ll c_{11}$, where
$\tilde{B}_{L,T}$ are defined in (\protect{\ref{lesb}}).
$\xi$ is the crossover length
defined in (\protect{\ref{crosslength}}). When $x < \xi$ the ratio takes
the random manifold value $2\nu+1$. It decrease slowly to $1$ in the
asymptotic regime $x \gg \xi$.}
\end{figure}

\begin{figure}[tbh]
\plotfig{l}{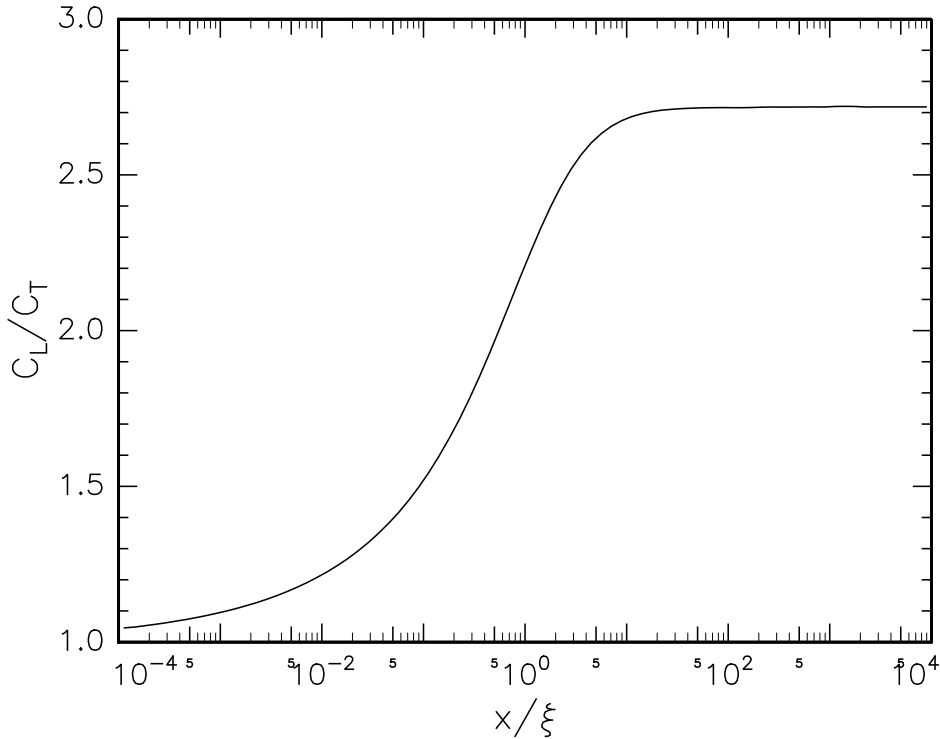}
\caption{\label{figure4}
Plot of the ratio of the translationnal order correlation functions
$C_{L}(r,z=0)/C_{T}(r,z=0)$ for the triangular Abrikosov
lattice, in the limit $c_{66} \ll c_{11}$.
$\tilde{C}_{L,T}$ are defined in (\protect{\ref{lesc}}).
$\xi$ is the crossover length
defined in (\protect{\ref{crosslength}}). The ratio increases from one
at short distance $x < \xi$ and saturates rapidly
to a universal number in the asymptotic regime
$x \gg \xi$. The variational method predicts this number to be $e$.}
\end{figure}

For the vortex lattice (\ref{total}), shear deformations dominate
($c_{66} \ll c_{11}$) in most of the phase diagram. The expressions
for the function $\tilde{B}^{L,T}(r)$ which describe the
crossover between the random manifold (intermediate distance)
regime and the large distance regime then simplify:
\begin{equation}
\tilde{B}^{L,T}(r)
\simeq
\frac{a^2}{\pi^2}\tilde{b}^{T,L}(\frac{r}{\xi})
\end{equation}
In the limit of weak disorder $\xi \gg a$ we find that there should
be a well defined crossover function, i.e all curves
should scale when plotted in units of $x/\xi$.
The relative displacement correlation functions $\tilde{B}^{T,L}$,
as predicted by the variational method, are
plotted in Fig.~\ref{figure1} and Fig.~\ref{figure2} for the triangular
lattice, by numerically integrating (\ref{simibani}).
The crossover
between the random manifold regime and the asymptotic quasi-ordered
asymptotic regime is apparent, and occurs at a scale of order $\xi$. At
the length scale $r = \xi$
where the random manifold regime cease to be valid,
the translationnal order correlation function $C_{K}(r) =
e^{-\frac12 K_\alpha K_\beta \tilde{B}_{\alpha,\beta}(r)}$, is of order
$C_{K_0} \sim 0.1$. Therefore the crossover should be
experimentally observable.
In Fig.~\ref{figure3}, we have shown the ratio $R$ of the transverse to
longitudinal displacements.
As was shown by BMY
\cite{bouchaud_variational_vortex_prl,bouchaud_variational_vortex},
its value is $2\nu + 1$ in the random manifold regime (the variational
method give $R=4/3$). At large scale, we find that this ratio decreases
to $R=1$, and in that sense isotropy is restored. However, if one
looks at the correlation functions for translational order, one finds
that the difference between
the longitudinal and transverse parts persists at large scales. Defining
the longitudinal and transverse translational correlation functions by
\begin{equation} \label{lesc}
C^{L,T}(r) = e^{-\frac{K_0^2}2 \tilde{B}^{L,T}(r)}
\end{equation}
$C^{L,T}(r)$ correspond to correlation functions with a
separation $r$,
parallel and perpendicular respectively to the vector $K_0$.
As is seen on Fig.~\ref{figure4}, the ratio $R_C=C^L/C^T$ increases from
one at short distances and saturates at a finite value at large
distances.
This value depends on the elastic constants, as seen from
(\ref{discor}). In the limit $c_{66} \ll c_{11}$, this number takes a
value which the variational method gives as athe universal
constant $e=2.7182..$.
The fact that $R_C$ saturates at large distance is a
consequence of the existence of the quasi-order. Had a random manifold
or a Larkin regime been valid up to large distances, this ratio would
increase indefinitely. On the other hand if the system was genuinely
ordered $R_C$ would saturate at a much smaller value than $e$, a value
which would go to one when $T\to 0$. As is discussed in more details in
section~\ref{experiments}, this should have observable experimental
consequences.

\subsection{Physical Discussion: crossover lengths, dislocations}
\label{discussion}

One can give a simple physical explanation for the three regimes found
here. Consider two flux lines separated in the ideal
lattice by $x$. In presence of disorder
the mean squared relative displacement is $\tilde{B}(x)$.
There is a length at which the potential seen by a line is smooth.
This length is the greatest of the correlation lengths of the random
potential
$\xi_0$, or the Lindemann length $l_T=\sqrt{\langle u^2\rangle_T}$. This
defines a separation between vortices which we have denoted by $l$ in this
paper
such that $\tilde{B}(l) \simeq \text{max}({\xi_0}^2 ,{l_T}^2)$.
At zero temperature it equal the length defined by
$R_c$ defined by Larkin Ovchinikov length ( and $L_c$ in the $z$ direction)
and in general $l$ can be thought of as the Larkin Ovchinikov length
renormalized
by temperature.
Below this length the elastic manifold
sees a smooth potential with well defined derivatives, thus
a local random force can be defined.
Indeed expanding in $u$ the disorder
potential energy in (\ref{disorder}) gives a random force term $f.u$
with
$f(x)= \sum_K V(x) K exp(-i K x)= \nabla V(R_i)$. In the sum over harmonics
the maximum $K$ is $K_{max}=2 \pi/\xi_0$. Thus
this expansion is valid only as long as $u \ll \xi_0$. This defines the
range of validity of the Larkin regime, i.e at $T=0$ $x < R_c$
and more generally $x<l$ ( we assume $R_c > a$).

For separations larger than $R_c$ but such that $\tilde{B}(x) \ll a^2$ each
flux line explores only its immediate vicinity and feels different disorder.
This is the regime explored by BMY which is identical to the random manifold.
This can be seen, on a more mathematical level,
from our model by summing over all the harmonics
for instance on the replicated Hamiltonian (\ref{cardyos}). One gets
$V(u) \sim \sum_{R_i} \delta(u_a - u_b - {R_i})$. For $u \ll a$ only the $R=0$
term contributes and each line sees an independent random potential.
This intermediate random manifold regime holds up to the length $x=\xi$
such that
$\tilde{B}(\xi) \approx a^2$ at which periodicity becomes important. There
is no gain in energy to shift the lattice by $a$. In this regime displacements
grow much more slowly and only the lowest harmonics contribute. This is
the quasi ordered regime.

In order to apply this theory to experimental systems one has in
principle to worry about topological defects, such as dislocations.
Although the influence of dislocations is still a controversial question,
their influence has been clearly overestimated in the past.
Let us mention some arguments, which we believe are incorrect,
put forward to argue that unbound dislocations will proliferate even
at weak disorder.
An Imry Ma type argument is the following.
The core energy cost of a dislocation cannot be avoided and scales as
$L^{d-2}$.
A dislocation loop of size $L$ creates extra-displacements of order O(1) up to
logarithms,
in a region of size $L^{d}$. By adjusting the position of the loop
one can hope to gain an energy from disorder $L^{d/2}$. Thus below $d=4$
dislocation will be favorable.
Such an argument is incorrect because it is again based on Larkin random force
model
for which disorder energy is linear in the displacement. For the real model
the energy varies as $\cos(2 \pi u/a)$ and adding a dislocation displacement
will not necessarily gain enough disorder energy.
In fact if the Larkin or the random manifold regime were true up to infinite
scales
it would indeed be favorable to create dislocations. The energy fluctuations
due to disorder is $L^{\theta} \sim L^{d-2+2 \nu} \gg L^{d-2}$.
If $\nu >0$ dislocations
will occur because it will always be energetically favorable to replace
an elastic
distortion by a dislocation \cite{Huse_private}. However in the case of a
lattice
or if quasi long range order is preserved in the system as it is the
case here, both energies scale the same way, since $\nu=0$, maybe up
to logarithms. The prefactor
of the disorder term can then be made arbitrarily small at weak disorder
while the core energy of the dislocation is a given finite number. Thus
if disorder is weak
enough it is likely that dislocations will not appear.
Even if they do the scale will be huge, and the effects associated with
disorder that we discuss in this paper should be observable
over a wide range of distances.
If disorder is gaussian one could also argue that rare fluctuations will
eventually
lead to dislocations at exponentially large scale. Realistic disorder however
is bounded and thus such effect should be absent.

\subsection{Experimental consequences} \label{experiments}

Let us now discuss in more details the experimental consequences of our
findings. Two main types of experiments exist at the moment to probe the
translational order of the vortex lattice: magnetic decoration experiments and
neutron diffraction experiments. One would expect for these two
experiments that the results of sections~\ref{theory}
and~\ref{discussion} apply. However, the direct comparison with
experiments could be complicated due to the effects
of the non-local elasticity and
3D anisotropy. These effects can be included, in principle, in the
variational
calculations by simply changing the elastic Hamiltonian, at the price of
extremely tedious calculations.
Even if
a detailed treatment of such effects is beyond the scope of this paper,
their importance can be estimated by the following simple arguments.

In the high-Tc
Abrikosov lattice, the elastic constants vary by orders of magnitude
when the wave-vector goes from $1/\xi_0$ to $1/\lambda$. A good
approximation of the elastic modulii for $H_{c1} \ll B \ll H_{c2}$ is
\cite{glazman_koshelev_decoupling,fisher_c44_calcul}
\begin{eqnarray} \label{nonlocal}
c_{44}(q) & = & \frac{B^2}{4 \pi}
               \frac{1}{1+\lambda^2 {q_z}^2 +{\lambda_c}^2
                 {q_\perp}^2 } + {c'}_{44} \nonumber \\
c_{11}(q) & = & \frac{1+{\lambda_c}^2 q^2}{1+\lambda^2 q^2} \\
c_{66}(q) & = & \frac{\Phi_0 B}{{(8 \pi \lambda)}^2}  \nonumber
\end{eqnarray}
where $\lambda$ is the London penetration depth in the $ab$ plane,
$\lambda_c=\Gamma \lambda$ along the $c$ axis with
$\Gamma=\sqrt{M_z/M}$
and $c'_{44}(q_z)$ is the single flux line contribution
to the tilt modulus. One must have also $B<H_{dec}=\Gamma^2 \Phi_0/d^2$
to avoid further effects of decoupling between planes, where $d$ is the
distance between $CuO$ planes.
Since in this regime $c_{66}$ is dispersionless and much smaller than
$c_{11}$, most of the effects of non local elasticity comes from the
$q_\perp$ momentum dependence of $c_{44}$.
In the region of Fourier space where $c_{44}$ varies
strongly, i.e
for $1 \ll \lambda_c {q_\perp} \ll \lambda_c a^2 $ a good approximation
to $c_{44}$ is
\begin{equation} \label{proper}
c_{44}\approx \frac{B^2}{4 \pi {\lambda_c}^2 {q_\perp}^2}
\end{equation}
this new momentum dependence of $c_{44}$ will lead to very different
lattice displacements. Using (\ref{relation}) with the proper $c_{44}(q_\perp)$
(\ref{proper}), one finds now $[\sigma](v)\sim v/\log^2(v)$. This leads,
using (\ref{btildedef}), to a very slow growth of the relative
displacements, $\overline{(\langle[u(x)-u(0)]\rangle)^2} \propto \log(x)$ or
constant, for $\lambda < x < \lambda_c$.
If the translational correlation length
$\xi > \lambda_c$, the random manifold regime will survive, whereas
if $\xi < \lambda_c$ one would expect the non local elasticity effects to
dominate the random manifold regime. Note however that the asymptotic
large distance regime, for $\xi > \lambda_c$ will be completely unchanged.

In decoration experiments however, one is usually in a regime of very
small fields. In particular $B < H_{c1}$ (bulk), giving $a > \lambda$.
For example, in the Bell experiments \cite{grier_decoration_manips}
performed on $10 \mu m$ thick samples of BSCCO,
one has $\lambda = 0.3 \mu \text{m}$ and
$\lambda_c \approx 60 \lambda$. The highest field picture (69 Gauss) have
very large regions free of dislocations, for which one can hope to apply
the theory of the present paper. For such fields $a \sim 2 \lambda$, and
thus $\lambda < a < \lambda_c$.
It is likely, since the
interactions between vortices are less important than in the regime $B
> H_{c1}$, that single vortex contributions will dominate. Thus the
$q_{\perp}$ dependence of $c_{44}$ will be weaker, and one can hope non
local elasticity to be unimportant.

Recently, we have carefully
reanalysed \cite{decoration_giamarchi_pld}
the data of Ref.\cite{grier_decoration_manips}, performing a
Delaunay triangulation of the larger field images which do not contain
dislocations.
This allows to compute $\tilde{B}(r)$ directly.
Preliminary results indicate a very good fit to a
power law behaviour $\tilde{B}(r) \sim r^{2 \nu}$ from
$r=a$ up to $r=30 a$ with an exponent $2 \nu =0.4 \pm .05$.
Thus, assuming dispersionless elastic constants, this
is a strong indication that one is seeing the random manifold regime.
The exact exponent $2 \nu$ for the random manifold regime
is unknown, but the (Flory) value $2 \nu=0.33$ predicted by the variational
method
is expected \cite{pld_machta_saw} to be a (relatively good) {\it lower bound}.
Another prediction \cite{feigelman_collective,halpin_frg}
for the exponent using refined scaling arguments,
which might turn out to be more accurate, is
$2 \nu= 4(4-d)/(8+m)$, i.e $2 \nu=0.40$ for $d=3$ and $m=2$.
The data exclude a Larkin type behaviour $\tilde{B}(x) \sim x$,
and in fact there seems to be
no measurable Larkin regime for small $r$, indicating that
$l \sim a$. One does clearly observe a saturation in $\tilde{B}(r)$
around $r=30-40 a$ at a value of $\tilde{B}(r)$ consistent with the predicted
saturation
to the slower logarithmic growth. However
larger pictures would be necessary to conclude unambiguously on
the crossover itself, as well as the large distance regime. The main
obstacle is of a statistical nature, i.e there are not enough pairs of
points uncorrelated statistically, to perform the necessary
translational average. Larger pictures would allow such an average to be taken.

Clearly both the understanding of the short distance regime, and
of the importance or not of non local elasticity, and the existence of
the quasi ordered regime deserve further studies.
Other difficulties in interpreting data from
decorations experiments can come from the fact that surface interactions
may be different from the bulk ones \cite{Huse_private}. It has been
argued recently however that the effects of the surface interactions may be
visible only at scales much larger than the size of
the decoration pictures \cite{marchetti_nelson_fits}.

Another good probe of the correlations in the vortex lattice, which is
free of potential surface problems, is the neutron scattering experiments.
Detailed neutron diffraction studies are now available for different
type II superconductors, such as NbSe$_2$
(see reference~\onlinecite{yaron_neutrons_vortex}) as well as BSCCO.
Neutron experiments measure (up to a form factor taking into account the field
distribution created by a single vortex line) the Fourier transform at $k = K_0
+ q$
of the density correlation around a reciprocal lattice vector $K_0$.
The structure factor which is measured is given by
\begin{equation} \label{neutrons}
S(q) = \int d^3 x e^{i q x}  e^{-\frac12 K_\alpha K_\beta
                     \tilde{B}_{\alpha\beta}(x)}
\end{equation}
where $\tilde{B}$ is given in (\ref{lesb} \ref{resdebut} \ref{simibani}
\ref{resfin}).
The full calculation of
$S(q)$ requires a numerical integration of (\ref{neutrons}), but
the main features can be given analytically. Let us recall that
$\xi$ is the translational
correlation length due to disorder defined in (\ref{crosslength})
and that trivial anisotropy has been taken into account by proper rescalings
of $z$ versus $r$ directions.

At small $q$, $q < 1/\xi$  the integral (\ref{neutrons}) is dominated
by the large
distance regime where $\tilde{B}(x) = A_3 \log(x)$, where $A_3 = 1$
according to the variational method. The structure factor
is therefore
\begin{equation} \label{divergence}
S(q) \sim (1/q)^{3-A_3}
\end{equation}
and thus {\it diverges} at small $q$, a consequence of the persistence of
quasi-long range order in the system. True Bragg peaks therefore exist.
This is in sharp contrast with
previous predictions assuming simple or stretched-exponential
decay of the translational correlation function up to large distance.
At a wavector of order $q \sim 1/\xi$, the
behavior of $S(q)$ will crossover to a slower decay, controlled by the
random manifold regime. In this regime
\begin{equation}
S(q) \sim (1/q)^{3+2 \nu}
\end{equation}
A clear signature that one is indeed in the regime (\ref{divergence})
described here should
show in the neutron experiments by the fact that $S(q)$ has no true half-width.
The maximum value of $S(q)$ will be limited
either by the experimental resolution of by the distance between
unpaired dislocations $\xi_D$. Varying these parameters
should leave the rest of the curve nearly
unchanged (this is valid as long as $\xi_D \gg \xi$).
The distance between dislocations could be controlled
by annealing the lattice, either using a driving force or
a field-cooling procedure similarly to what
is done in Ref. \onlinecite{yaron_neutrons_vortex}.

Another interesting prediction can be made for the in-plane,
$q_\perp$-dependent ratio $S(q_\perp \parallel K_0)/S(q_\perp \perp K_0)$.
After integration over $q_z$ the structure
factor becomes, at small $q_\perp$
\begin{equation}
S(q_\perp,z=0) = \int dq_z S(q_\perp,q_z) \sim \frac{2\pi}{q_\perp} e^{\frac12
-
                                       (\hat{K_0}\cdot\hat{r})^2}
\end{equation}
Thus the ratio goes for small $q_\perp$ to the value
\begin{equation}
\frac{S(q_\perp \perp K_0,z=0)}{S(q_\perp\parallel K_0,z=0)} = e
\end{equation}

\newpage

\section{Functional Renormalization Group} \label{degal4}

Another method widely used to study disordered problems
is the functional renormalization group. It turns out that its
application to the present problem due to the periodicity in
(\ref{cardyos}) is simple. It provided a good complement to the
variational method, none of the methods being rigorous. The functional
RG can only give results in an $\epsilon= 4-d$ expansions, which does
not have presently the status of rigor of the standard $\epsilon$ expansions of
usual critical phenomena for pure systems.
In particular the effects of multiple
minima will affect higher orders in perturbation theory
and could very well result in replica symmetry breaking instability
in the FRG flow, as found recently in Ref. \cite{ledoussal_rsb_lettre} (see
also
Section \ref{degal2}). On the other
hand the functional RG
should include fluctuations more accurately than the variational
method, provided it does not miss another part of the physics.
Comparison of the two methods near four dimensions should allow to test
their accuracy.

\subsection{One component model}

For simplicity we confine our study to a model with isotropic elasticity
as in (\ref{cardyos}). Let us first consider
$u$ to be a scalar field ($m=1$) and set $c=1$. The full replicated
Hamiltonian is
\begin{equation}
H/T = \frac{1}{2T} \int d^dx (\nabla u(x))^2
- \frac1{2T^2} \sum_{ab} \int d^dx \Delta(u_a(x) - u_b(x))
\end{equation}
For simplicity we take $K_0 =2 \pi$, so that
the function $\Delta(z)$ is periodic of period $1$. In the original
Hamiltonian $\Delta(z)$ is a sum of cosine given in (\ref{cardyos}).
One then performs the standard rescaling $x\to e^l x$ and $u\to
e^{\zeta l}u$. The idea of this renormalization is to perform an
expansion around a classical solution at zero temperature. One should
keep the whole function $\Delta$ in the renormalization procedure.
The RG equations to order $\epsilon = 4-d$ have been derived by D.S. Fisher
\cite{fisher_functional_rg} for the random manifold problem (see also
\cite{halpin_frg,balents_frg_largen})\cite{natterman_leschhorn}
\begin{eqnarray} \label{scalarfrg}
\frac{d \Delta}{dl} & = & (\epsilon - 4 \zeta) \Delta  + \zeta z \Delta'
+ \frac12 (\Delta'')^2 - \Delta'' \Delta''(0) \\
\frac{d T}{dl} & = & (2-d) T \nonumber
\end{eqnarray}
A factor $1/S_d=2^{d-1}\pi^{d/2}\Gamma[d/2]$
has been absorbed into $\Delta$ in (\ref{scalarfrg}).
The temperature is an irrelevant variable and flows to zero.
The correlation function
\begin{equation}
\tilde{\Gamma}(q) = T \tilde{G}(q) = \langle u^a(q)u^a(-q) \rangle
\end{equation}
satisfies the RG flow equation
\begin{equation}  \label{rgflowfrg}
\tilde{\Gamma}(q,T,\Delta) =
e^{(d+2\zeta)l} \tilde{\Gamma}(q e^l,T e^{(2-d)l}, \Delta(l))
\end{equation}

The periodicity of the function $\Delta$
implies that the roughening exponent
is $\zeta=0$ for the large distance behavior.
This allows us to obtain the only periodic fixed point
function $\Delta^*(z)$ in the interval $[0,1]$:
\begin{equation}  \label{fixedpoint}
\Delta^*(z) = \frac{\epsilon}{72} (\frac1{36} -z^2(1-z)^2)
\end{equation}
Values for other $z$ are obtained by periodicity. The fixed point is
stable except for a constant shift, which corresponds to a change in the
free energy. The linearized spectrum is discrete
and the eigenvectors can be obtained using hypergeometric functions.
The fixed point function is non analytic
at the origin. It has a singular part which
behaves as $z^2|z|$, for small $z$ and leads to $\Delta^{*(4)}(0) = \infty$.
As discussed below this is a general feature of fixed points for this
type of disordered systems \cite{fisher_functional_rg}.
For a periodic fixed point, i.e. $\zeta=0$, one can set
$e^{l^*} q = 1/a$ in (\ref{rgflowfrg}) this
allows to obtain perturbatively, provided $l^* \gg 1$
\begin{equation} \label{rgflow2}
\tilde{\Gamma}(q,T,\Delta) = \left(\frac1{qa}\right)^d
               \tilde{\Gamma}(\frac1a,T \simeq 0,\Delta^*)
\end{equation}
One can evaluate the correlation in (\ref{rgflow2}), at a scale of the
order of the cutoff, perturbatively in $\Delta$. One then gets
\begin{equation} \label{rgflow3}
\tilde{\Gamma}(q,T,\Delta) = \left(\frac1{qa}\right)^d
 \left. \frac{-\Delta''^*(0)}{k^4}\right|_{k=1/a}
\end{equation}
Using (\ref{fixedpoint}), and remembering the factor $1/S_d$ in
$\Delta$, one obtains
\begin{equation}
\tilde{\Gamma}(q) = \frac{a^{4-d}\epsilon}{36 S_d q^d}
\end{equation}
This gives
\begin{equation} \label{decayfrg}
\overline{\langle (u(x) - u(0))^2 \rangle}=\frac{\epsilon}{18} \log|x|
\end{equation}
whereas the variational method gives
\begin{equation}
\overline{\langle (u(x) - u(0))^2 \rangle}=\frac{\epsilon}{2\pi^2}
\log|x|
\end{equation}
(\ref{decayfrg}) gives a power law decay for the translational
correlation functions with an exponent
$A_{d,RG} = \epsilon (2 \pi)^2/36=1.10\epsilon$ against
$A_{d,RG} = \epsilon$ for the variational method.
The agreement of the two methods on the exponent $A_d$ is within
10\%, which is very satisfactory. The fact that $A_{d,VAR} \leq
A_{d,RG}$ is not surprising since the variational method
underestimates a
priori the effect of fluctuations. One can remark that omitting the term
$(\Delta'')^2/2$ in (\ref{scalarfrg}) leads to a fixed point
$\Delta^*(z) = \cos(2\pi z)/(2\pi)^2$, which gives exactly the same
exponent than the variational method.

At intermediate distance it is enough to focus on the small $u$
behavior of the function $\Delta$, and thus to forget in effect the
periodicity. At short distances the function $\Delta$ is analytic. In
that case \cite{fisher_functional_rg}
\begin{equation}
\frac{d \Delta''(0)}{d l} = (\epsilon-2\zeta)\Delta''(0)
\end{equation}
Setting $\zeta=\epsilon/2$ allows to get a fixed point $\Delta(z) =
A z^2 - 2 A^2$. Using
(\ref{rgflowfrg}) one obtains
\begin{equation}
\tilde{\Gamma}(q) \sim \frac{\Delta''(0)}{q^4}
\end{equation}
which corresponds to the Larkin random force regime. This however holds
only at short scales. This fixed point is unstable, and a non
analyticity at $z=0$ develops, corresponding to an algebraic decay of
the $\Delta_K$ in (\ref{cardyos}). $\Delta$ eventually
flows towards the long distance regime described by the fixed point
(\ref{fixedpoint}). There might be an intermediate random manifold
regime.

Another renormalization method that has been used, was a real space RG
by Villain and Fernandez
\cite{villain_cosine_realrg}. For $2<d<4$ this method, which is
approximate, also predicts a logarithmic growth of the correlations.
It does not allow however to compute the universal prefactor $A_d$ or
the crossover function. The agreement between these methods,
none being rigorous, lends credibility to
the additional results in $d=3$ obtained using the variational method.

\subsection{General case}

Let us consider now the more general case of
an $m$ component vector $u$, and isotropic
elasticity (\ref{iso}). The equation
giving the fixed point function becomes instead of (\ref{scalarfrg})
\begin{equation} \label{fullfrg}
\Delta[u] + \frac12(\partial_\alpha\partial_\beta \Delta[u])^2
     - \partial_\alpha\partial_\beta \Delta[u=0]
       \partial_\alpha\partial_\beta \Delta[u] = 0
\end{equation}
while the displacement correlation function becomes,similarly to
(\ref{rgflow3})
\begin{equation} \label{fullamplitude}
\overline{\langle u_\alpha(q) u_\beta(-q) \rangle}  =
        (\partial_\alpha\partial_\beta \Delta^*[u=0]) \frac1{q^d}
\end{equation}
where $\epsilon$ has been included in $\Delta$.
For the case $m=1$, (\ref{fullfrg}) reduces to (\ref{scalarfrg}). For
the case $m=2$, the analysis depends on the symmetry of the lattice.
For a square lattice a separable function
\begin{equation}
\Delta[u_x,u_y] = \Delta^*[u_x] + \Delta^*[u_y]
\end{equation}
where $\Delta^*$ is a solution of (\ref{scalarfrg}) satisfies (\ref{fullfrg})
and gives
(\ref{meansq}) with the same exponent $A_{d,RG}$ than for the case
$m=1$. A rectangular lattice would give the same exponent.
The triangular lattice is more difficult to treat and no simple
solution of (\ref{fullfrg}) can be found. We have performed a numerical
solution of (\ref{fullfrg}). There is a non trivial solution which has
the full symmetry of the triangular lattice. In that case one would
expect in general different exponent $A_d$ than for the square lattice,
unless there is an
hidden symmetry reason, for which the exponent does not depend on the
lattice symmetry.
It is difficult to get a high precision for the exponent because of the
non analytic nature of the solution. The numerical value found for $A_d$
was within $5\%$ of the one for the square lattice, but we were unable
to decide within our accuracy whether the two exponents were equal or
different. The exponent is
again very close from the one predicted by the variational method, which
is independent of the lattice symmetry. Once again, neglecting the term
$\frac12(\partial_\alpha\partial_\beta \Delta[u])^2$ in (\ref{fullfrg}), one
recovers exactly the result of the variational method.

\newpage

\section{d=2} \label{degal2}

In $d=2$, thermal fluctuations are expected to play a more important
role. Already in the case
of the pure system, they change the true long-range order of the lattice into
a power law decay of the correlation functions, with an exponent
controlled by the temperature. One can therefore expect a stronger
competition between disorder and temperature than in higher dimension.
In addition standard renormalization group techniques are available in
$d=2$ and can be compared with the variational method.
In the section~\ref{theoryd2} we examine
the $d=2$ problem using both the variational method and the
renormalization group. We will focus mainly on $d=1+1$
(flux lines in a plane). The results are mostly relevant there
since the starting model (\ref{cardyos})
becomes exact, due to the fact that dislocations cannot exist in $d=1+1$.
The physical consequences for various experimental systems both in
$d=1+1$ and $d=2+0$ will be discussed in section~\ref{experimentd2},
together with the effects of dislocations.

\subsection{Theoretical results} \label{theoryd2}

In $d=2$ the variational method applied to the starting model
(\ref{cardyos}) leads to a solution
which belongs to the class of
``one-step'' replica symmetry breaking
\cite{giamarchi_vortex_short,korshunov_variational_short},
in some extended sense, i.e such that
$[\sigma](v)=0$ vanishes for $v<v_c$ and $[\sigma](v) >0$
for $v_c<v<1$. This can be seen readily
by taking the limit $d \rightarrow 2^+$
in (\ref{sigmeq}), a limit which vanishes
for $v<v_0=T K_0^2/(4 \pi c)$. This solution represents a glass
phase. Since $v_0$ cannot be larger
than $1$, the glass phase exists in $d=2$ for $T<T_c=4 \pi c/K_0^2$.
For $T > T_c$ the disorder is irrelevant, and the replica symmetric solution
is stable, as already
discussed in section~\ref{repsym}. The detailed behavior below $T_c$
will again depend
on whether one considers simply a single cosine model, or takes into account
all
the harmonics present in (\ref{cardyos}).
For simplicity we will focus here on the single cosine
model, which has been simulated numerically and is interesting in itself.
The main effect of the higher harmonics is again to allow for a
random manifold crossover regime, at low enough temperature
in the glass phase. It is examined in details
in appendix~\ref{random2}.

In the case of the single cosine model (\ref{singlecosine})
one can look simply for a constant solution $[\sigma](v) =
\Sigma_1$ for $v > v_c$.
The details of the calculations can be found in appendix~\ref{stab},
where the saddle point equations (\ref{saddlebreak})
for $\Sigma_1$ and $v_c$ are derived. These equations are solved in
(\ref{circular}) to give
\begin{eqnarray} \label{circulart}
v_c & = & \frac{T K_0^2}{4\pi c} \frac{\Lambda}{\Sigma_1}\log[1+
                   \frac{\Sigma_1}{\Lambda}]  \\
\Sigma_1 & = & K_0^2 \frac{\Delta v_c}{m T} \left(\frac{\Lambda}{\Sigma_1}
               + 1 \right)^{-K_0^2 T/(4 \pi c)}
\nonumber
\end{eqnarray}
where the cutoff $\Lambda \simeq c (2 \pi/a)^2 $
Assuming the cutoff to be very large (i.e $\xi \ll a$) in (\ref{circulart})
one gets
\begin{eqnarray} \label{vc}
v_c & = &  \frac{T K_0^2}{4\pi c} \\
\Sigma_1 & = & \Lambda \left(\frac{\Delta K_0^4}{4 \pi m
                   \Lambda}\right)^{1/(1-T/T_c)}
\nonumber
\end{eqnarray}

As we will show below $\Sigma_1$ defines a characteristic length
\begin{equation} \label{corrlength}
\xi(T) = \sqrt{c/\Sigma_1} =
a(c^2 a^2/\Delta)^{1/(2-T K_0^2/2 \pi c)} =
a(c^2 a^2/\Delta)^{1/(2 - 2 T/T_c)}
\end{equation}
above which there is
a crossover to the asymptotic regime dominated by the disorder.
The above expression for $\xi(T)$ in $d=2$ is valid for $\xi \gg a$.
At zero temperature it
coincides with the length found
using simple Fukuyama Lee arguments (see section~\ref{dimensionarg})
and it is renormalized
downwards by thermal fluctuations at finite temperature, an effect
specific to two dimensions.

Using
(\ref{tilde}) and the form (\ref{gtildeap}) for $\tilde{G}(q)$ one
obtains for the relative displacement
\begin{eqnarray} \label{integj}
\tilde{B}(x) & = & \frac1m \overline{\langle (u(x) - u(0))^2 \rangle }
 \\
    & = &  \tilde{B}_0(x)+ 2T(\frac1{v_c} -1)\frac1{(2\pi)}
           \int_0^{\infty}
           k dk (\frac1{c k^2} - \frac1{c k^2+\Sigma_1})(1-J_0(k x))
\nonumber
\end{eqnarray}
where $\tilde{B}_0(x)$ is the value of $\tilde{B}$ in the absence of
disorder
\begin{equation}
\tilde{B}_0(x) = \frac{T}{c \pi}\log(\Lambda x)
\end{equation}
(\ref{integj}) is convergent although each term is individually
divergent, but is easily regularized by multiplying by $J_0(\epsilon k)$
with $\epsilon\to 0$. This leads to
\begin{equation} \label{fullb}
\tilde{B}(x) = \frac{T}{c \pi}\log(\Lambda x) + \frac{T_c-T}{c \pi}
\left(\log(x/\xi) + K_0(x/\xi) + \gamma + \log(1/2) \right)
\end{equation}
where $\gamma$ is the Euler constant. The expression (\ref{fullb}) gives
the following crossover for $\tilde{B}(x)$
\begin{eqnarray} \label{asyb}
\tilde{B}(x) & \simeq & \frac{T}{c \pi} \log(x) \qquad x \ll \xi \\
\tilde{B}(x) & \simeq & \frac{T_c}{c \pi} \log(x) \qquad x \gg \xi
\nonumber
\end{eqnarray}
where $T_c = 4 \pi c / K_0^2$. Note that for $a \ll x \ll \xi$ there is
in principle a
Larkin regime where one has algebraic growth of the disorder part of
the correlation function with $2 \nu_L = (4-d) =2$
plus logarithmic corrections:
\begin{equation}  \label{larkinsimp}
\tilde{B}(x) = \tilde{B}_0(x) + \frac{T_c-T}{4 c \pi}
( \log(\xi/(2 x)) + 1 - \gamma) (\frac{x}{\xi})^2
\end{equation}
However, except at very low temperatures, the thermal part
always exceed the disorder part and thus disorder effects are masked
by thermal effects at short distances.

The variational method predicts therefore a simple logarithmic growth of
the displacements at large distances,
both above $T_c$ and below. The effects of disorder are
limited to the freezing of the prefactor to the value $T_c$ for
temperatures below $T_c$. Note that $T_c$ is a universal quantity,
independent of the strength of the disorder. Such a result is valid in
the limit where the ultra-voilet cutoff $\Lambda$ is very large.
The disorder strength enters
the crossover length $\xi$ above which the asymptotic behavior for $T <
T_c$ can be observed. Of course $\xi \to \infty$ when $T \to T_c$
as can be seen from (\ref{corrlength}).

Note that the effect of the cutoff, which could be important
for a numerical simulation not at small disorder, lead to
some temperature dependence of the amplitude of the logarithm.
The amplitude in (\ref{asyb}) becomes
\begin{equation}
A=T/(v_c c \pi)=T_c/(c \pi) \frac{\Sigma_1/\Lambda}{\log(1+\Sigma_1/\Lambda)}
\end{equation}
where one can use $\Sigma_1$ from (\ref{vc}) or better from the
equation (\ref{circular2}) for $y$ in the appendix~\ref{stab}.
One finds an increase of the amplitude $A$ when the temperature
decreases.

In $d=1+1$, it is also possible to write renormalization group equations
for the disorder
\cite{villain_cosine_realrg,cardy_desordre_rg,toner_log_2}.
To lowest order, and on a square lattice
for simplicity, such RG equations were derived by Cardy and Ostlund
\cite{cardy_desordre_rg} and read:
\begin{equation} \label{rg1}
\frac{d\Delta}{dl} = (2- \frac{K_0^2 T}{2 \pi c}) \Delta - C \Delta^2
\end{equation}
where we denote $\Delta=\Delta_{K_0}$, and
\begin{equation} \label{rg2}
\frac{d\Delta_0}{dl} = \frac{{K_0}^2 a^4}{ T^2} \Delta^2
\end{equation}
Both $\Delta$ and $\Delta_0$ are defined in (\ref{cardyos}). $C$ is a
constant \cite{ledoussal_rsb_lettre} which is unimportant for our
purposes. For $T > T_c$ the disorder is irrelevant, in agreement with
the variational method. For $T < T_c$ the disorder term becomes
relevant and there
is a new non-trivial fixed point at $\Delta=(T_c-T)$. This fixed point
however has the unusual feature that the variable $\Delta_0$ flows to
infinity  $\Delta_0(l) \propto l $.
However since this variable does not feed back at any order
in perturbation theory (only averages of the type
$\overline{(\sum_a C_a \phi_a)^2}$ with $\sum_a C_a=0$ appear) it has been
assumed that this fixed point was correct.

Using the RG one can again define a short and large distance regime. At
short distances the RG is certainly correct, and is
more accurate than the variational method since it
treats the fluctuations correctly.
As was noted in section~\ref{singlecos}, at short scales $x$ such
that $|u(x)-u(0)| \ll \xi$ (for the single cosine model),
it is possible to expand the cosine to equivalently recover
the Larkin random force model. The correlation function for that
model reads simply
\begin{equation} \label{simply}
\tilde{G}(q) \propto  \frac{1}{q^2} + \frac{\Delta}{q^4}
\end{equation}
This usually leads to $\delta \tilde{B}(x) \sim x^2$, but here
one must take into account renormalization by thermal
fluctuations. This is done by integrating the RG equation in the
small distance regime where we note that
(\ref{rg1}) is correct as long as $\Delta a^2/T^2 \ll 1$, which
is equivalent to $(a/\xi)^2/(l_T/a)^2 \ll 1$. One can integrate (\ref{rg1})
to obtain
\begin{equation}
\Delta(l) = \Delta(0) e^{l (2- 2\frac{T}{T_c})}
\end{equation}
Applying the RG flow equation
\begin{equation} \label{rgflow}
\tilde{G}(q,T,\Delta_0,\Delta) =
e^{2l} \tilde{G}(q e^l,T, \Delta_0(l),\Delta(l))
\end{equation}
where $\tilde{G}$ has been defined in (\ref{tilde}), immediately leads to
\begin{equation}
\delta \tilde{G}(q) \sim \frac{\Delta(l^*)}{q^2}
\sim \frac{1}{q^{(4-2T/T_c)}}
\end{equation}
The RG therefore predicts that the
Larkin regime is in fact {\it anomalous} with
an exponent continuously varying as a function
of the temperature
\begin{equation} \label{anom}
\delta \tilde{B}(x) = \tilde{B}(x)-\tilde{B}_0 (x)
\sim a^2 ( x/\xi )^{2 - \frac{T K_0^2}{2 \pi c}}
\end{equation}
instead of (\ref{larkinsimp}),
for $x<\xi$ (in the low-T regime $\xi$ is replaced by another length).
There are also corrections coming from the renormalization of
$\Delta_0$. Integrating (\ref{rg2}) one gets
\begin{equation}
\Delta_0(l)=\Delta_0(0) +
\frac{K_0^2 a^4 \Delta(0)^2}{2 T^2 (2- \frac{T}{T_c})}
(e^{2 l (2- \frac{T}{T_c})}-1)
\end{equation}
but such corrections are obviously smaller at short distances.

At large distance $\Delta_0(l) \gg 1$, in order to obtain the correlation
functions, one has to assume that the unusual CO fixed point is indeed
correct. If one does so,
correlation functions
can be computed \cite{toner_log_2}
using RG flow equation (\ref{rgflow}).
Iterating until $l^*$ such that
$e^{l^*} q = 1/a$ allows to obtain for large $l$:
\begin{equation} \label{rgflowbis}
\tilde{G}(q,T,\Delta_0,\Delta) =
\frac{1}{q^2}  \tilde{G}(a,T,\Delta_0(l),\Delta^*)
\sim  \log(1/q) /q^2
\end{equation}
which leads to $\tilde{B}(x) \sim \log^2(x)$.
In (\ref{rgflowbis}) it
has been assumed that simple perturbation theory could be done for
the correlation functions at scale $x=a$.
The RG approach would
therefore predict a $\log^2(x)$ growth of the displacements.
at variance with the predictions of the variational method
which gives a simple log. In fact the RG results is based on the
assumption of replica symmetry. As we have shown
recently\cite{ledoussal_rsb_lettre},
a careful analysis of the Cardy Ostlund
fixed point and of the RG flow shows that it is unstable to
replica symmetry breaking.
When RSB is allowed one obtains a runaway flow of the RG which is consistent
with the findings of the variational method.
Two recent numerical calculations on this model
\cite{batrouni_numerical_cardy,cule_numerical_cardy} seem to confirm
that the GVM does describe the correct physics at large distance.
None of them is
compatible with a $\log^2(x)$ growth of the displacements.
In Ref. \onlinecite{batrouni_numerical_cardy}, no change in the static
correlation functions
was observed in the presence of small disorder, whereas a transition
occurring in the dynamic correlation functions was observed at $T_c$. A
careful comparison was performed with the predictions of the RG
calculations, and the results were found incompatible. These numerical results
are
however consistent with the prediction of the variational method.
Indeed for such weak disorder the length $\xi(T)$ is very large,
especially near $T_c$, and
simulations performed on a too small system will show no deviations
as is obvious from (\ref{asyb}).
However in Ref. \onlinecite{cule_numerical_cardy} the disorder is much
larger, and $\xi(T=0)$ is of the order of the lattice spacing $a$. This
simulation indeed shows quite clearly a freeze of the amplitude of the
logarithm below $T_c$ at the value $A=T_c/(c \pi)$.

\subsection{Physical realization in $2+0$ dimensions: Magnetic bubbles}
\label{experimentd2}

Elastic models in two dimensions have been studied for some time.
The Hamiltonian (\ref{cardyos})
describes several physical disordered systems,
such as randomly pinned flux arrays in a plane
\cite{fisher_vortexglass_short,nattermann_flux_creep,giamarchi_vortex_short},
the surface of
crystals with quenched bulk or substrate disorder
\cite{toner_log_2}, planar Josephson junctions
\cite{vinokur_josephson_short},
domain walls in incommensurate solids. In addition,
a very nice realization, investigated in detail recently,
is provided by magnetic bubbles \cite{seshadri_bubbles_long}.
In such systems one should be able to
test the predictions of the previous section. However, if the elastic
objects are not lines ($d=1+1$) but points ($d=2+0$) it is now important
a priori to take into account the effects of free dislocations.
In two dimensions,
dislocations are expected to be much more important than in three
dimensional systems, and to have observable consequences on the
destruction of translational and/or orientational order.
A very naive argument
in $d=2$ is that while the energy of a dislocation pair of separation $r$
increases as $\sim T_0 Log (r/a)$ in the absence of disorder (a slowly
growing function
but still allowing for a quasi-solid phase for $T<T_0$), in the
presence of disorder, the energy should saturate
to a finite value $E_{\text{max}} \simeq T_0 Log (\xi/a)$ when $r > \xi$.
This is
because long range order is supposed to be destroyed beyond this length,
effectively screening the elastic pair interaction.
If it is the case, unpaired dislocations
will be thermally excited at any finite
temperature with weight $e^{-E_{max}/T}$.
Note that the same type of argument in $d=3$ excludes dislocation loops
of size $r$ whose energy scale as $r Log (\xi/a)$ and
thus cannot proliferate so
easily. This was discussed in section~\ref{discussion}.

As was discussed in section~\ref{theoryd2}, in the absence of
dislocations quasi-long range order persists in the system even in the
presence of disorder, and the
above naive arguments need to be reexamined.
A quantitative theory including both disorder and dislocations is
difficult. One can however get an idea of the effects of dislocations by
using the renormalization equations of Cardy and Ostlund in the presence
of topological defects \cite{cardy_desordre_rg}.
These equations were derived by Cardy and Ostlund for a
one component field (XY model) with both disorder (random field) and vortices.
They correspond to the modification of (\ref{rg1}) and (\ref{rg2}) to
include the dislocation fugacity, and an additional equation for the
renormalization of the dislocation fugacity itself \cite{cardy_desordre_rg}.
Generalizing them to a triangular lattice beyond linear order goes beyond the
scope
of this paper, but we do not expect radically new conclusions, rather
a change by a small factor of the parameters ( temperature, etc..)
(note that extension to a $n=2$ components on a square lattice is
obvious). There are several reasons why we believe that
the CO equations might {\it overestimate} the effect of dislocations.
One is that, as we saw before (\ref{theoryd2})
these RG equations implicitely assume replica
symmetry. They lead to a faster decay of the translational correlation function
in the absence of dislocations, of the type $\text{exp}(-\log^2(r))$
and thus overestimate the effect of disorder compared to the GVM,
which includes replica symmetry breaking.
The GVM predicts that quasi long range order exists in the
system. Thus the CO equations should also overestimate the effects of
dislocations.

A stability diagram can be constructed by examining the
renormalization equations to linear order, for the fugacity of topological
defects and for the disorder, In the system
considered by Cardy and Ostlund \cite{cardy_desordre_rg}
the fugacity of vortices $y$ satisfies to lowest order in y:
\begin{equation}\label{fugacity1}
\frac{dy}{dl} = 2 y~(1 - \frac{\pi c}{2 K_0^2 T} +
\frac{ \pi {\rho_0}^2 \Delta_0}{2 K_0^2 T^2} )
\end{equation}
while the $q=K_0$ component of the disorder $\Delta_{K_0}$
renormalizes to linear order as:
\begin{equation} \label{fugacity2}
\frac{d\Delta_{K_0}}{dl} = 2 \Delta_{K0} (1 - \frac{K_0^2 T}{4 \pi c})
\end{equation}

The influence of {\it internal} disorder on a two dimensional
crystal was studied in Ref. \cite{nelson_elastic_disorder}.
In that work the disorder
was taken as a quenched random stress coupling to the strain, i.e
a term $- \sigma(x) \nabla \cdot u$ was added to the hamiltonian
with $\overline{\sigma(x) \sigma(x')} = {\rho_0}^2 \Delta_0 \delta(x-x')$.
As discussed in \ref{derivation} this corresponds to include only
the {\it long wavelength} part of the disorder, and to neglect the
$q=K_0$ component of the disorder, which is very important for the
present problem of {\it substrate} disorder. For the
2D elastic triangular lattice the fugacity of dislocations $y$
renormalizes, to linear order, in a way very similar to
\ref{fugacity1}:
\begin{equation}\label{fugacity3}
\frac{dy}{dl} = 2 y~(1 - \frac{a^2}{4 \pi T} \frac{c_{66}
(c_{11} - c_{66})}{ c_{11} }
+ \frac{\rho_0^2 \Delta_0 a^2}{\pi T^2} {(\frac{c_{66}}{c_{11}})}^2 )
\end{equation}
One easily sees that, to linear order,
the $q=K_0$ disorder renormalizes also very similarly to \ref{fugacity2}
\begin{equation} \label{fugacity4}
\frac{d\Delta_{K_0}}{dl} = 2 \Delta_{K_0}~ (1 - \frac{K_0^2
(c_{11}+c_{66}) T}{8 \pi c_{11} c_{66}})
\end{equation}

\begin{figure}[tbh]
\plotfig{r}{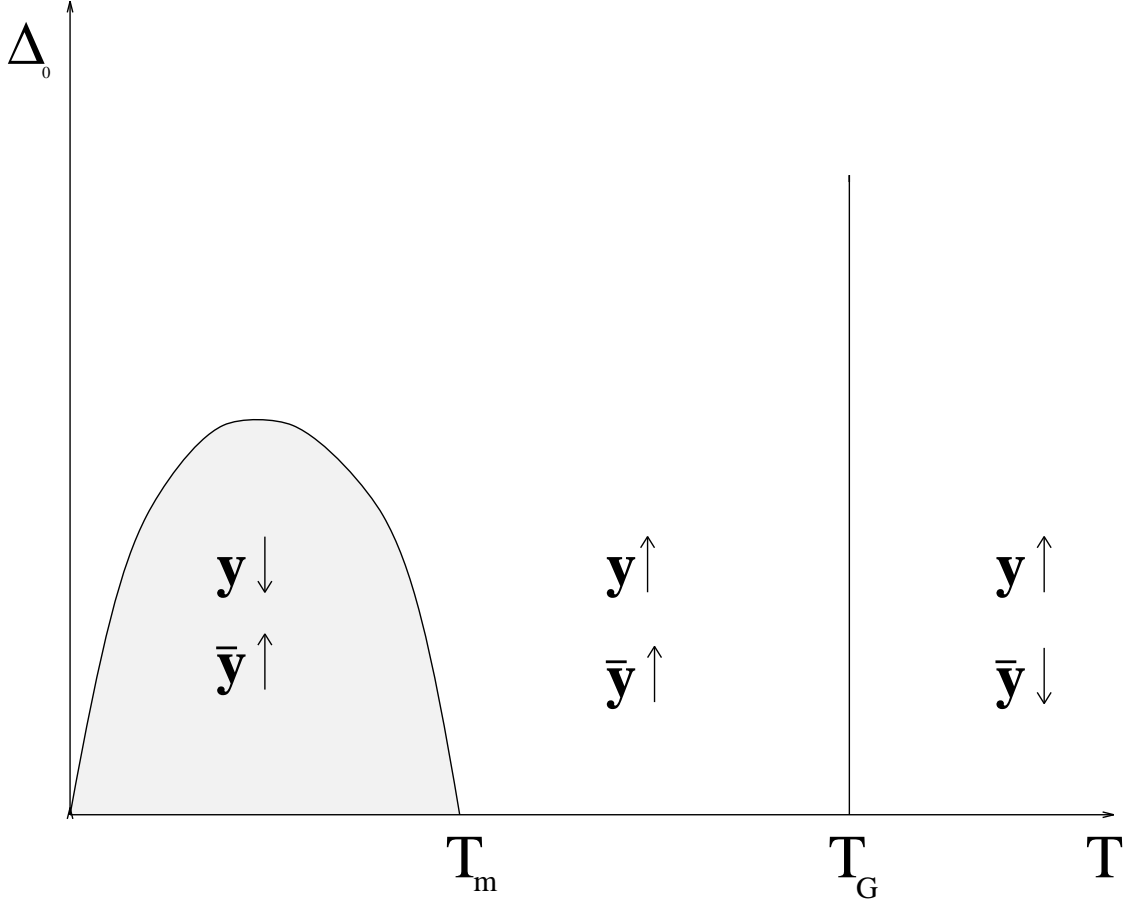}
\caption{\label{figdis}
Stability diagram of a two dimensional solid with weak quenched substrate
disorder, as a function of the temperature $T$ and the long wavelength
(i.e harmonic) part of the disorder $\Delta_0$. The
diagram indicates schematically the stability of the harmonic part of the
free energy to dislocations (fugacity $y$) and short-wavelength disorder
($\overline{y} \sim \Delta_{K_0}$) by showing the
relevance of these variables to linear order.}
\end{figure}

The stability diagram is shown in figure~\ref{figdis}.
In presence of dislocations there are thus various possible
regimes. There are two critical temperatures. One is the
KTHNY melting temperature $T_{m}=a^2 c_{66}
(c_{11} - c_{66}) / 4 \pi c_{11}$, above which dislocations
unbind for the pure system which melt into a hexatic.
The other one is the glass transition temperature
$T_G=8 \pi c_{11} c_{66} / K_0^2 (c_{11}+c_{66})$,
below which disorder becomes relevant in
the absence of dislocations. The ratio between these
temperatures is equal to half of the exponent $\eta_{K_0}$ of
the pure system at $T_m$. For the CO model it is thus
universal and equal to $T_{m}/T_G = 1/8$. For the triangular
lattice it depends on the (renormalized) elastic constants
\cite{nelson_halperin_melting}
but cannot exceed $1/6$. In figure~\ref{figdis}
we have shown schematically how $y$ and $\overline{y} \sim \Delta_{K_0}$ are
renormalized, at linear order as a function of temperature
and long wavelength disorder $\Delta_0$ from \ref{fugacity3} \ref{fugacity4}.
At high temperature dislocations are relevant and disorder
is washed away by thermal fluctuations. Below $T_G$
disorder becomes relevant leading, in the absence
of dislocations, to the Cardy-Ostlund line of fixed points. This line
however is {\it unstable} to dislocations. In the shaded region in
figure~\ref{figdis}, the dislocations would be perturbatively irrelevant for
the
system with $\Delta_{K_0}=0$ and the solid would survive
( in fact in that case if one includes non linearities in $y$ the
solid is stable in an even greater region \cite{nelson_elastic_disorder} ).
However in this region the disorder
$\overline{y} \sim \Delta_{K_0}$ is relevant. It will eventually increase
$\Delta_0$ as can be seen from (\ref{rg2}), and drive the flow towards
the region where $y$ itself increases and unpaired dislocations
will eventually appear at large scale.

Let us now estimate at which scale dislocations will appear, assuming
weak disorder. The
disorder $\Delta_{K_0}=0$ becomes of order one at lengthscale of order
$\xi$. The key point is that
up to this lengthscale the fugacity of dislocation has been
renormalized {\it downwards} and is now much smaller with
$y(\xi)/y(a) \sim (a/\xi)^{2(1-T_{m}/T)}$.
At this lengthscale, one ends up with a system
for which the disorder is of order one and the fugacity of dislocations
is extremely small. One can therefore predict that the typical distance between
unpaired dislocation $\xi_D$ in the shaded region is much larger than $\xi$,
the
distance above which effects of disorder manifest and (\ref{asyb}) can be
observed. It is impossible to compute rigorously the ratio $\xi_D/\xi$
using perturbative RG since beyond $\xi$ one of the coupling constant
($\Delta_{K_0}$)
is large, but one can still estimate this ratio by the following
heuristic argument. If one assumes that, when $\Delta_{K_0}$ has become of
order
one, (\ref{rg2}) is still valid, then above length of order $\xi$,
\begin{equation}
\Delta_0(l) = \Delta_0 + \alpha l
\end{equation}
where $\alpha$, is a coefficient of order one. Then estimating the
distance between dislocations as the scale at which $y$ becomes of order
one, one gets using \ref{fugacity1}
\begin{equation}
\xi_D \sim \xi \exp{\left( C \log^{1/2}((\xi/a)^{2(T_{m}/T -1)}/y(a)) \right) }
\end{equation}
where $C$ is a constant of order unity and $y(a)$ the bare value
of the dislocation fugacity.

The resulting prediction is that in the shaded region of
figure~\ref{figdis} dislocations appear only at scales
large compared to $\xi$, and the main reason for decay of
translational correlations is the pinning of the elastic
manifold. Thus the elastic theory
developped in this paper should apply up to distances up to order
$\xi_D$. In that low temperature region perturbative RG
does not allow to obtain precisely the
correlation functions even at not too large distance
since one is far from the perturbative region $T \sim T_G$.
The variational method on the other hand predicts
the following results: as shown in appendix~\ref{random2}, there is a
crossover temperature $T^*=T_G/\log(\xi/a)$. For temperatures
$T>T^*$ there is no intermediate random manifold regime and
the displacement correlation function should show
the logarithmic growth of (\ref{asyb}). However since the disordered
solid regime corresponds to $T < T_{m} \simeq T_G /8$, one is likely to be in
the regime $T < T^*$. In that case there is a random manifold regime at
short distances where the displacement correlation function grows as
$\tilde{B}(x) \sim (x/\xi)^{2/3}$. At large distance one recovers the
logarithmic asymptotic regime. The anisotropy ratio between the
longitudinal and transverse displacements $R=\tilde{B}_T/\tilde{B}_L$
should crossover from $2\nu+1\simeq 1.7$ in the random manifold regime
to $1$ in the asymptotic regime.

Experimentally, in the system of magnetic bubbles
\cite{seshadri_bubbles,seshadri_bubbles_long} some regimes
are found indeed where the distance between dislocations $\xi_D$ is of the
order of up to five times the translational correlation length.
The authors of Ref. \cite{seshadri_bubbles_long} observe a crossover between a
regime where $\xi_D \sim \xi$ to a regime where $\xi_D \gg \xi$, which
probably corresponds to the transition to the shaded region discussed above.
Indeed, the measured ratio $\xi_D/\xi$ keeps increasing with density
and is limited there only by experimental setting.
It would be very interesting to increase the range experimentally
accessible.
In Ref. \onlinecite{seshadri_bubbles_long}
some comparison with the random manifold result was also performed.
However the definition of the
anisotropy ratio used in Ref. \onlinecite{seshadri_bubbles_long}
($\xi_{T}/\xi_{L}$)
does not seem appropriate, since
the data were first fitted to a simple Lorentzian shape. New insight in
the physics of such disordered 2D systems could be gained by a
reanalysis of these data. More theoretical work is also needed
since the
above analysis of the dislocations is very crude. It
has been
generally assumed that the non perturbative RG flow in presence of dislocations
was towards a kind of hexatic, but nothing is really known on this
phase. In particular,
in the Cardy-Ostlund approach, the vortices are supposed to
be completely thermalized and to see only a
uniform potential, and feel no effect of the disorder. This is
physically incorrect, since in a disordered system the fugacity will also
depend on the
position. One should therefore take also into account
terms which will pin the vortices. This is another reason
why we believe that the CO analysis overestimates the effect of
topological defects.

To conclude, $d=2$ elastic disordered systems have therefore a very rich
behavior.
Many other experimental systems could be investigated, among them
colloids \cite{murray_colloid_hexatic,murray_colloid_prb} and thin
superconductors.
A realization which would exclude dislocations could
be to adsorb a polymerized membrane, with very high
dislocation core energy, on a disordered substrate.

\newpage

\section{Conclusion} \label{conclusion}

In this paper we have developed a quantitative description of the static
properties of a lattice in presence of weak disorder. We have derived
a model (\ref{cardyos})
valid in the elastic limit $\nabla u \sim a/\xi \ll 1$
which contains the proper physics
at all lengthscales, and allows to describe the various regimes as function
of distance.
We have applied the Gaussian Variational Method (GVM)
to this model and computed
the correlation functions of the relative displacements.
This method has the advantage of being
applicable in any dimension.
The comparison with the renormalization group study that we
have also performed, which is possible in
$d=4-\epsilon$ and in $d=2$, indicates that the GVM is an accurate
variational ansatz for this problem. It also shows that the
GVM should be a good tool to explore other disordered problems
where different length scales are present.
Contrary to previous
studies the present analysis includes both the effect of metastable
states and of the intrinsic periodicity of the lattice, both effects
being found to be very important.

We found that the effect of impurities on the translational order
of the lattice is weaker than was previously thought and that quasi-long range
order
persists at large scales. The resulting phase, which we call a ``Bragg
glass'' has both the properties of being a glass and of being
quasi-ordered. The analogy with the quasi-order which subsist
at low temperature in pure two dimensional solids is puzzling.
The Bragg glass possess several intrinsic length scales. At
very short scales it behaves as predicted by Larkin's random force
model. This regime should be quite limited in $d=3$ at low temperature
and for a potential rough at scales smaller than the lattice.
In $d=2$ however, thermal effects being stronger, this regime
is wider. At intermediate scales, and low temperature in $d=2$,
the system behaves as a random manifold of flux lines independently
pinned by impurities with stretched exponential decay of translational order.
At distances larger than $\xi$, we have shown that due to the periodicity
of the lattice,
the pinning by impurities becomes less effective and quasi order survives.

We have not attempted to include explicitly topological defects
in the present quantitative study.
However, we have given evidence both in $d=3$ and $d=2$
that for weak impurity disorder
the effects of dislocations is less important than is usually
believed. This is due to the fact that
quasi-long range order subsists even in
the presence of disorder for the elastic system.
In $d=2$ it is possible, but by no means established, that
unpaired dislocations will appear at large scale.
We have estimated conservatively the length scale at which this happens and
found that it can be much larger than $\xi$. Thus in $d=2$
there is a regime where the decay of translational order
is due primarily to the elastic displacements induced by disorder.
In $d=3$, we have argued that there is probably a phase
{\it without unpaired dislocations} for weak disorder
and at low temperature.

Thus
in the three-dimensional high temperature superconductors
there could be generically be (at least) {\it two} types of
glass phases caused by point impurities, i.e
two types of vortex glasses. The first one is the
strong disorder vortex glass phase which
is described by the gauge glass type of models
\cite{fisher_vortexglass_short,fisher_vortexglass_long}
and presumably contains a lot of dislocations. The second
one is the quasi-ordered Bragg glass
described in the present paper.
A natural speculation then is that the tricritical point recently
observed in experiments
\cite{safar_tricritical_prl}, whose position in the $H-T$ plane
can be raised by adding more point impurities in a controlled fashion
\cite{kwok_electron_defects}, has something to do with the
relevance of dislocations. It could mark the
separation between these two phases since it is
natural that the quasi-ordered glass melts through
a sharper transition \cite{charalambous_melting_rc},
much like a pure solid, while
the vortex glass transition of Ref.
\cite{fisher_vortexglass_short,fisher_vortexglass_long}
is believed to be continuous.
\cite{fisher_vortexglass_short,fisher_vortexglass_long}
In fact, it is not even clear whether, strictly speaking,
the strongly disordered vortex glass phase of
Ref. \cite{fisher_vortexglass_short,fisher_vortexglass_long}
is a true thermodynamic phase, rather than a very long crossover
from the flux liquid. Indeed recent simulations
indicate that $d=3$ could be slightly
below its the lower critical dimension \cite{bokil_young_vglass}.
The Bragg glass, on the other hand, should not suffer from such
existential problems.

The results of the present study suggest some comments concerning experiments.
First
decoration experiments should be reanalyzed. Clearly a fit to a simple
exponential as inspired by Ref.
\cite{chudnovsky_pinning,chudnovsky_pinning_long}
is certainly inadequate for pictures where no dislocation is present.
Since we predict that the crossover to the asymptotic quasi-ordered
regime occurs when the correlation function $C_{K_0}(r)$ is of order $0.1$
this crossover should be observable, in principle.
One must keep in mind that Abrikosov lattices might not be the ideal
system to compare with the theory, since they are very complicated
objects with additional intrinsic scales.
In particular the effect of nonlocal elasticity, 2d to 3d crossover effects
and surface interactions, although they can in principle be incorporated into
the method, complicate the analysis by introducing
new crossover lengths such as the
London length. It would be therefore highly desirable, in order to test
the above predictions to investigate simpler elastic systems. One example
are colloids which have been very successful to investigate translational order
in pure systems \cite{murray_colloid_hexatic,murray_colloid_prb}.
In addition,
these experiments could help decide whether dislocations are a thermodynamic
property of the state or simply a non equilibrium feature which can be
eliminated.
Neutron diffraction experiments could also help decide on these issues.
We obtained several characteristic predictions for neutron diffraction,
such as the existence of algebraically diverging Bragg peaks
and predictions for the ratio of transverse to longitudinal scattering
intensities.

We have addressed here only the statics of a lattice in presence
of disorder, the quantities of interest in that case being
mainly related to positional order. It would be clearly very interesting to
obtain information on the {\it dynamics} and especially
the driven dynamics of such systems. This is indeed particularly
important for superconductors where most experiments
measure only dynamical quantities. Several remarks
are in order.

There exist presently a qualitative approach to
the dynamics of the vortex glass based mostly on scaling arguments
for energy barriers \cite{feigelman_collective}.
On such a qualitative level one usually assumes
a single scale for the energy landscape, i.e that barriers
$E_b(L) \sim L^\theta$ scale with the same exponent $\theta$
which describes energy fluctuations in the statics.
Assuming that there is a regime of transport where plastic
deformations can be neglected, i.e that energy barriers are
controlled only by elastic motion, the present study of the
statics can be used to describe the creep regime at low
temperature. The argument goes that
in presence of an applied external force $f$, such as
the Lorentz force created by an external current $|f| \sim |j|$,
the barrier for an optimal deformation $u \sim L^{\nu}$
away from a low energy configuration
on a scale $L$ is lowered and becomes
$E_b(L) \sim L^\theta - f u L^d$.
To unpin locally the manifold thus needs to go
over an Arrhenius barrier
$E_b \sim 1/f^\mu$ with $\mu=\theta/(d+\nu-\theta)$. This, and the exponents
found here
lead to a non linear voltage-current relation
$V \sim \exp[-1/(Tj^{\mu})]$ in the flux creep
regime, where $\mu$ crosses
over from $\mu \approx 0.7-0.8$ to $\mu=1/2$ as $j$ decreases.

On a more fundamental level one anticipates that the
methods used here for the statics and the results
they yielded will set the ground for the study of the dynamics.
As was discovered recently \cite{kurchan_dynamics}
the hierarchical structure
of states which is encoded in the replica symmetry breaking solution
of the statics finds an exact translation in the dynamics.
There it corresponds to the existence of a
long time dynamics for which the fluctuation dissipation theorem
breaks down and non equal time correlation and response functions
become highly non trivial. It would be interesting to
generalize these studies to the flux lattice. Let us finally note that
the quantities computed here using the statics are
equal time disorder-averages and they will coincide with the observed
translational averages in a given experiment provided
equilibrium has been reached, which will always hold below
a certain length scale.

We thank D.J. Bishop, J.P. Bouchaud, M. Charalambous, D. Fisher,
M. Gabay, D. Huse, T. Hwa, M. Mezard, C.M. Murray, R. Sheshadri.
for interesting discussions

\appendix
\section{Relabeling of the lines} \label{density}

In this appendix we detail the derivation of expression
(\ref{transp}) for the decomposition of the density
in term of the relabeling field (\ref{slowly}). We denote
$d$-dimensional positions $x=(r,z)$ where $r$ belongs to the
$m$-dimensional transverse space.
The density is given by
(\ref{densistart})
\begin{equation} \label{depart}
\rho(x) = \sum_i \delta(r - R_i -u_i(z))
\end{equation}
In order to take the continuum limit, one can introduce a smooth
displacement field $u(r,z)$ by
\begin{equation} \label{smooth}
u(r,z) = \int_{BZ} \frac{d^dm}{(2\pi)^m} e^{i q_\perp r} \sum_j e^{-i q_\perp
R_j}
          u_j(z)
\end{equation}
such that $u(R_i,z)=u_i(z)$
which has no Fourier components outside of the Brillouin zone (BZ).
In terms of the smooth field (\ref{smooth}) one can introduce the
relabeling field
\begin{equation} \label{definition}
\phi(r,z) = r - u(\phi(r,z),z)
\end{equation}
In the absence of dislocations there is a unique solution of
(\ref{definition}) giving $u(r,z)$ as a function of $\phi(r,z)$. $\phi$
is a m-component smooth vector field labeling the lines, and which takes
an integer-like values at
each location of a line
\begin{equation}
\phi(R_i + u(R_i,z),z) = R_i
\end{equation}
Substituting (\ref{definition}) in (\ref{depart}) one gets
\begin{equation} \label{tempo}
\rho(x) = \sum_i \delta(R_i - \phi(r,z))
          \text{det}[\partial_\alpha \phi_\beta(r,z)]
\end{equation}
Using the integral representation of the $\delta$ function,
(\ref{tempo}) becomes
\begin{equation} \label{perfectla}
\rho(x) =  \text{det}[\partial_\alpha \phi_\beta] \int
           \frac{d^dq}{(2\pi)^d} \rho_0(q) e^{i q \phi(x)}
\end{equation}
where
\begin{equation} \label{mdensity}
\rho_0(q) = \sum_i e^{i q R_i}
\end{equation}
For the case of a perfect lattice $\rho_0(q)$ is
\begin{equation} \label{sdensity}
\rho_0(q) = \rho_0 (2\pi)^d \sum_K \delta(q-K)
\end{equation}
Using (\ref{sdensity}) in (\ref{perfectla}) one gets formula
(\ref{transp}).
\begin{equation} \label{transpa}
\rho(x) =  \rho_0 \text{det}[\partial_\alpha \phi_\beta] \sum_K
e^{i{K}\cdot\phi(x)}
\end{equation}
Assuming that we are in the elastic limit
$\partial_\alpha u_\beta \ll 1$ one can expand (\ref{transpa}) to get
\begin{equation} \label{finaldens}
\rho(x) \simeq \rho_0 [1 - \partial_\alpha u_\alpha (\phi(x)) +
        \sum_{K \ne 0} e^{i K (r - u(\phi(x),z))}]
\end{equation}
In (\ref{finaldens}) one can replace $u(\phi(r,z),z)$ by $u(r,z)$ up to
terms of order $\partial_\alpha u_\beta \ll 1$. Note that in doing so
$u$ has negligible (suppressed by powers of $a/\xi$) Fourier components
outside the Brillouin zone, and thus there is a complete decoupling
between the gradient term and higher $K$ terms.

The same procedure can be carried on in the case
where the equilibrium lattice of the $R_i$ contains topological
defects such as dislocations, vacancies etc. Suppose for instance
that one wants to study a lattice with a fixed number of dislocations
at prescribed positions in the internal coordinate of the lattice,
i.e a network with a fixed topology (connectivity), but which is
now allowed to fluctuate in the embedding space due to coupling to
disorder and thermal noise.
This is relevant for the physical situation of a flux lattice
with {\it quenched-in } dislocations whenever one can neglect
dislocation motion, i.e glide and climb. The equilibrium positions
$R_i$ now correspond to a minimum of the elastic energy with
the constraint of prescribed connectivity. The problem at hand is to
analyze the extra small elastic displacements around the
equilibrium position due to disorder.
The density is still given by (\ref{perfectla})
where $\rho_0(q)$ is now the Fourier transform of the lattice
of the $R_i$ (\ref{mdensity}).

Upon coupling to disorder, the equivalent of the last
term in (\ref{cardyos}) will be generated. After averaging
over the random potential it reads:
\begin{equation}
\int \frac{d^dq}{(2\pi)^d} \frac{\Delta_q}{2T}
\cos(q (u^{a}(x)-u^{b}(x))) S(q)
\end{equation}
where $S(q)=\rho_0(q)\rho_0(-q)$ (with no averages) is the structure factor of
the
lattice without disorder, with fixed connectivity and
in its equilibrium position.
If in addition one wants to allow for topological defects to equilibrate
one needs to perform some further average over connectivities,
which is a difficult task.

\section{Link with quantum models in $1+1$ dimensions}
\label{quantuma}

The variational method can also be applied to study disordered one
dimensional (space) interacting bosons or fermions.
To see that fact
we use a representation of operators in terms of phase fields
introduced by Haldane \cite{haldane_bosons}. This representation
maps the system into an elastic hamiltonian similar to the
one used to describe flux lattices in \ref{derivation}.

The single-particle
creation operator is written:
\begin{equation} \label{boso}
\Psi^\dagger(x) = [\rho(x)]^{\frac12} e^{i\theta(x)}
\end{equation}
where $\rho(x)$ is the particle density operator and $\theta(x)$
the phase of the $\Psi$ field. To take into account the discrete nature
of the particle density, one introduces an operator $\Phi$
which increases by $\pi$ at each particle's location. The density
operator then is
\begin{equation}\label{devel}
\rho(x) = \frac1\pi\frac{\partial\Phi(x)}{\partial x}
\sum_{m=-\infty}^{+\infty}\exp[2im\Phi(x)]
\end{equation}
$\Phi$ can be expressed in term of another operator $\phi$ by $\Phi =
\pi\rho_0 x + \phi$, and $\rho_0$ is the average
density. The $\phi$ and $\theta$ fields obey the canonical commutation
relations
\begin{equation}
[\phi(x),\frac1\pi \theta(x')] = i\delta(x-x')
\end{equation}
Using (\ref{boso}), the single particle operator becomes\cite{haldane_bosons}
\begin{equation} \label{singlepart}
\Psi^\dagger(x) = [\rho_0 + \frac1\pi \nabla \phi(x)]^{1/2}
                  \sum_m e^{i m (\pi\rho_0 x + \phi(x))} e^{i \theta(x)}
\end{equation}
For fermions the sum over $m$ in (\ref{singlepart}) is only over odd
$m$, whereas for boson the sum is only over even values of $m$.
For fermions $\pi\rho_0$
can be replaced in (\ref{singlepart}) by $k_F$, where $k_F$ is the
Fermi momentum.
The long-wavelength -- low-energy properties of the
interacting boson or fermion gas are described
by the Hamiltonian \cite{haldane_bosons}:
\begin{equation} \label{hambosd}
H = \frac{1}{2\pi}\int dx\,\{(\frac{v}{K})
(\partial_x \phi)^2 +  (v K)
(\partial_x\theta)^2\}
\end{equation}
When going to the Lagrangian, one gets
\begin{equation}
{\cal L} = \frac1{2\pi K} \int dx d\tau [\frac1{u} (\partial_\tau
\phi)^2 + u (\partial_x \phi)^2]
\end{equation}
which is obviously the same than the classical Hamiltonian
(\ref{iso}).

 From Galilean invariance one has $(v K)/\pi = \rho_0/m$,
and $v^2 = 1/(\kappa \rho _0 m)$, where $\kappa$
is the compressibility. $v$ and $K$ are therefore functions of the
interactions and incorporate all the interaction effects.
The excited states of $H$ are sound waves with phase velocity $v$.
For the
fermion problem the noninteracting case corresponds to $v=v_F$ and
$K=1$. For repulsive interactions $K<1$, whereas $K>1$ for attractive
ones. For the boson problem $K\to \infty$ when the repulsion between
bosons goes to zero, and $K$ decreases for increasing repulsive
interactions. For the case of a $\delta$ function repulsion, $K$ varies
from $\infty$ to $1$. $K=1$ would correspond to an infinite on site
repulsion. $K<1$ can be obtained only if longer range interactions are
considered.
The coefficient $K$ determines the asymptotic behavior of the
correlation functions\cite{haldane_bosons}. For the bosons one gets
\begin{equation}  \label{corfo}
\begin{array}{lcl}
\langle\Psi^\dagger_B(r)\Psi_B(0)\rangle
& = & B\rho_0(\rho_0r)^{-1/(2 K_b)} \\
\langle\rho(r)\rho(0)\rangle & = & 2 K_b (2\pi\rho_0 r)^{-2} +
A\rho_0^2(\rho_0r)^{-2 K_b}\cos(2\pi\rho_0r)
\end{array}
\end{equation}
with some numerical constants $A$ and $B$.
whereas the fermion correlation functions are
\begin{equation}
\begin{array}{lcl}
\langle\Psi^\dagger_F(r)\Psi_F(0)\rangle
&  = & B\rho_0(\rho_0r)^{-(K_f+K_f^{-1})/2} e^{i \pi \rho_0 r} \\
\langle\rho(r)\rho(0)\rangle & = & 2 K_f (2\pi\rho_0 r)^{-2} +
A\rho_0^2(\rho_0r)^{-2 K_f}\cos(2 k_F r)
\end{array}
\end{equation}
up to an angular part.
The coupling to a random potential uncorrelated in
space and time  will give again a Lagrangian
identical to (\ref{cardyos}). Terms of the form
\begin{equation}
{\cal L}_{\text{dis}} = \sum_p \sum_{a \ne b} \int dx d\tau D_p
          \cos(2p(\phi^a(x,\tau) - \phi^b(x,\tau)))
\end{equation}
where $p$ is an integer and $D_p$ some constants proportional to the
disorder will appear. They correspond to Fourier
components of the random potential close to $2\pi p \rho_0$.
The Fourier components close to $q=0$ correspond to forward scattering
and give terms similar to the gradient terms in (\ref{cardyos}).

As discussed in Section \ref{degal2} space and time uncorrelated disorder
is relevant when $K \le 1$. The transition point $K=1$ corresponds to
non-interacting fermions or equivalently to bosons interacting with
infinite $\delta$ repulsive potential. Disorder is thus
always relevant for fermions with repulsive interactions, and
always irrelevant for fermions with attractive interactions.
Disorder is irrelevant for bosons with only finite $delta$-function repulsion,
while it becomes relevant for bosons with sufficiently
strong longer range interactions.
One experimental realization corresponds to flux lines
in superconductors, confined to a plane, which can be realized
by proper alignement of the magnetic field. It was argued in
Ref \cite{nattermann_flux_creep} that this always lead to a glass phase, i.e
$K<1$.
This does not seem correct to us. If the vortex line
interaction was the sum of an infinite repulsive delta (i.e forbidden
crossings) and an extra repulsive interaction of finite range $\lambda$,
then disorder would indeed always be relevant. But this is not the
case, and the on-site interaction is finite, which
presumably leaves room for both a glass phase ($K <1 $)and a high temperature
phase ($K > 1$) in this experimental system.

\section{One-step solution for $d \le 2$} \label{stab}

In this section we examine a one step replica symmetry broken solution.
For simplicity we look at the model with a single harmonic
(\ref{singlecosine})
We now search a one step symmetry breaking solution of the form
\begin{eqnarray} \label{onestep}
\sigma(v) & = &  \sigma_0 \qquad [\sigma](v) = 0 \qquad v < v_c \\
\sigma(v) & = & \sigma_1 \qquad [\sigma](v) = v_c(\sigma_1-\sigma_0)
= \Sigma_1
\qquad v > v_c \nonumber
\end{eqnarray}
With the form (\ref{onestep}) one gets
\begin{eqnarray}  \label{gtildeap}
\tilde{G}(q)  & = &  \frac{G_c(q)}{v_c} + (1 - \frac1{v_c})
\frac{1}{G_c^{-1}(q)+\Sigma_1} + \sigma_0 G_c^2(q) \\
G(q,v) & = & \frac{\Sigma_1 G_c(q)}{v_c(G_c^{-1}(q) + \Sigma_1)}
                          +     \sigma_0 G_c^2(q) \qquad v > v_c \\
G(q,v) & = & \sigma_0 G_c^2(q) \qquad v < v_c \\
\ [G]  & = & 0 \qquad v < v_c \\
\ [G]  & = & \frac{\Sigma_1 G_c(q)}{(G_c^{-1}(q) + \Sigma_1)}
                               \qquad v > v_c
\end{eqnarray}
To determine $v_c$ one has to minimize the free energy
\begin{eqnarray} \label{libre}
F/(n m \Omega) & = & \frac{T}2 \int \frac{d^dq}{(2\pi)^d}
\biggl[ c q^2 \tilde{G}(q) - \log(G_c(q)) \nonumber \\
 & & - \frac{G(q,0)}{G_c(q)}
     + \int_0^1 \frac{dv}{v^2} \log\left[\frac{
        G_c(q) -  [G]}{ G_c(q)} \right] \biggr] \\
 & & + \frac{\Delta}{2 m T} \int_0^1 dv e^{-\frac{K_0^2}2 B(0,v)}
\nonumber
\end{eqnarray}
where
\begin{eqnarray} \label{bzerovc}
B(0,v) & = & 2 T \int \frac{d^dq}{(2\pi)^d}
\frac{1}{G_c^{-1}(q) + \Sigma_1} \qquad v \ge v_c \\
B(0,v) & = & B(0,v \ge v_c) + 2 T \int \frac{d^dq}{(2\pi)^d}
\frac{\Sigma_1}{G_c^{-1}(q)(G_c^{-1}(q) + \Sigma_1)v_c} \qquad v < v_c
\nonumber
\end{eqnarray}
To get the saddle point equations one has to differentiate
(\ref{libre}) with respect to $\sigma_0(q)$, $G_c$, $\Sigma_1$ and
$v_c$. Differentiation with respect of $\sigma_0(q)$ gives $G_c^{-1}(q)
= c q^2$. Differentiating with respect of $1/v_c$ and $\Sigma_1$ and
then replacing $G_c^{-1}(q)$ by its value one gets
\begin{eqnarray} \label{saddlebreak}
\Sigma_1 & = & \frac{K_0^2 \Delta v_c}{m T} e^{- K_0^2 T \int
               \frac{d^dq}{(2\pi)^d} \frac1{c q^2 + \Sigma_1}} \\
\Sigma_1 v_c & = & T K_0^2 \int \frac{d^dq}{(2\pi)^d}
     \left[\log(1 + \frac{\Sigma_1}{c q^2}) - \frac{\Sigma_1}{c q^2
           + \Sigma_1}\right]   \nonumber
\end{eqnarray}
Note that $\sigma_0$ does not appear in these equations. One could in
principle determine the value of $\sigma_0$ by differentiating
(\ref{libre}) with respect to $G_c^{-1}(q)$. However it is much easier
to use the saddle point equations (\ref{saddle}). For $d \leq 2$
one has
\begin{eqnarray} \label{pac}
\sigma_1 & = & K_0^2 \frac{\Delta}{m T} e^{-K_0^2 T \int \frac{d^dq}{(2\pi)^d}
\frac{1}{c q^2 + \Sigma_1}}\\
\sigma_0 & = & \sigma_1 e^{-K_0^2 T \int \frac{d^dq}{(2\pi)^d}
\frac{\Sigma_1}{c q^2(c q^2 + \Sigma_1)v_c}}
\end{eqnarray}
Thus $\sigma_0=0$ for $d \leq 2$. Note that in order to determine
$v_c$ one should not substitute in (\ref{libre}) the expressions
(\ref{pac}) {\it before} differentiating with respect to $\Sigma_1$ and
$1/v_c$. Such expressions are only valid at the saddle point.

In $d=2$ the above equations read:
\begin{eqnarray} \label{circular}
\Sigma_1 v_c & = & T K_0^2 \frac{\Lambda}{4\pi c}\log[1+
                   \frac{\Sigma_1}{\Lambda}]  \\
\Sigma_1 & = & K_0^2 \frac{\Delta v_c}{m T} \left(\frac{\Lambda}{\Sigma_1}
               + 1 \right)^{-K_0^2 T/(4 \pi c)}
\nonumber
\end{eqnarray}
where we used a circular cutoff $\Lambda \simeq c (2 \pi/a )^2$. Let us denote
by $T_c=4 \pi c/K_0^2$ and $y=\Sigma_1/\Lambda$. We then have:
\begin{eqnarray} \label{circular2}
v_c & = & \frac{T}{T_c} \frac{\log(1+y)}{y}  \\
y^{(1-T/T_c)} & = & \frac{K_0^2 \Delta}{m T_c \Lambda}
\frac{\log[1+y]}{y(1+y)^{T/T_c}}
\nonumber
\end{eqnarray}
We are interested in the case $ \frac{K_0^2 \Delta}{m T_c \Lambda}<1 $ since
$\xi \gg a$ and thus there is clearly a one step solution
for $T<T_c$. Note that for arbitrary $\Lambda$
$\Sigma_1$ vanishes when $T \to T_c^{-}$ and simultaneously $v_c$ goes to $1$.

For $d < 2$ the above equations lead to
\begin{eqnarray}\label{complicated2}
v_c & = & T K_0^2 (\frac{2 - d}{d}) j_d \Sigma_1^{\frac{d-2}2} c^{-d/2}\\\
\Sigma_1 & = & \frac{K_0^2 \Delta v_c}{m T} e^{- K_0^2 T
               j_d \Sigma_1^{(d-2)/2} c^{-d/2}}
\end{eqnarray}
where
\begin{equation}
j_d=\int \frac{d^dq}{(2\pi)^d} \frac{1}{q^2+1}=
\frac{\pi^{1-d/2}}{2^d \sin(\pi d/2) \Gamma[d/2]}
\end{equation}

We can compare the free energy (\ref{libre}) of the one step solution
with the free energy of the symmetric one,
that one obtains by
taking $v_c=1$ and $\Sigma_1=0$. The free energy difference $\Delta F =
F_{\text{onestep}} - F_{\text{rs}}$ is
\begin{equation} \label{complicated}
\frac{\Delta F}{m n \Omega}
= \frac{T}2 (\frac{1}{v_c}-1) \int \frac{d^dq}{(2\pi)^d} \left[
\frac{\Sigma_1}{c q^2 + \Sigma_1} - \log(\frac{c q^2 +\Sigma_1}{c
q^2}) \right] + (1-v_c) \frac{\Delta}{2 m T} e^{-K_0^2 B(0,v_c)/2}
\end{equation}
where $B(0,v_c)$ is obtained from (\ref{bzerovc}).
Using the saddle point equations (\ref{saddlebreak}),
(\ref{complicated}) simplifies into
\begin{eqnarray}
\frac{\Delta F}{m n \Omega} = \frac{\Sigma_1}{2K_0^2} \frac{(1-v_c)^2}{v_c}
\end{eqnarray}
Thus whenever there exist a one step solution it has higher energy than
the replica symmetric solution.

Examining the equations \ref{complicated2} one sees that indeed a
one step solution survives in $d=1$ even while the replica symmetric solution
is stable there. Note however that the one step solution changes nature
and, as an approximation to the original problem, acquire some unphysical
features. In $d=1$ it does not exist for small disorder at fixed $T$.
When it exist the transition
is discontinuous and $\Sigma_1$ jumps from zero to some non zero value. The
equations
are in fact very similar to a variational approach to roughening in $d=1$
where it is obvious that the cosine term is irrelevant and the
variational approach cannot be trusted beyond weak disorder, where
it gives sensible results. Here it is also clear that
$q=K_0$ disorder (i.e the cosine term) is irrelevant at large scale for $d<2$,
in the sense that it will not pin the $d=1$ manifold. It will however
renormalize the long-wavelength disorder which simply adds to the thermal
roughening $u \sim L^{1/2}$. The physical meaning
of the one step solution for $d=1$ being unclear, although it could be
related to transitions in the dynamics, we will not considering further.
Note finally that in $d=0$ the one step solution
does not exist at all ( since the mass is zero).

\section{Variational calculation on the non-local model}
\label{bmy}

If one does not use the decomposition (\ref{transparent}) of the
density, the variational equation for the off diagonal self energy is
\cite{bouchaud_variational_vortex_prl,bouchaud_variational_vortex}
instead of (\ref{bebe}) (in the presence of the extra disorder
(\ref{unphysical}))
\begin{equation} \label{variational}
\sigma_{a \ne b}(q_\perp)  =  \frac{\Delta}{T} \sum_i \frac1{R_i^\lambda}
\frac1{B_{ab}^{m/2+1}(R_i)}
(\frac{m}2 - \frac{R_i^2}{2B_{ab}(R_i)})
e^{-\frac{R_i^2}{2B_{ab}(R_i)}} \cos(q_\perp R_i)
\end{equation}
As for the local model (\ref{cardyos}) one has to look for a non replica
symmetric solution. Let us use the Poisson formula valid for arbitrary
lattice of $R_i$:
\begin{equation} \label{poisson}
vol(R_i) \sum_{R_i} \delta(x-R_i) = \sum_{K} e^{i K x}
\end{equation}
where $vol(R_i)$ is the volume of the unit cell and the $K$ are the
reciprocal lattice vectors. One can then replace the discrete sum in
(\ref{variational}) by an integral and insert the
integration over $x$ in formula (\ref{variational}). One gets
\begin{equation} \label{offdiag}
\sigma(q_\perp,v) = \frac{\rho_0 \Delta}{T} \sum_K \int d^m r e^{i K r}
\frac1{r^\lambda} \frac1{B^{m/2+1}(r,v)}
(\frac{m}2 - \frac{r^2}{2 B(r,v)})
e^{-\frac{r^2}{2B(r,v)}} \cos(q_\perp r)
\end{equation}
 From this exact expression, BMY kept only the components with $K=0$.
As pointed out by BMY,
\cite{bouchaud_variational_vortex_prl,bouchaud_variational_vortex}
if one takes $v$ small enough, then $B(r=1,v)$ is large
enough so that, for all the $r$ contributing to
(\ref{offdiag}), one can replace $B(r,v)$ by $B(0,v)$.
The idea is that $B(r,v)$ varies much more slowly that $r^2$, which is
true in the elastic limit.
Indeed one can easily see that for small $v$ and large $r$ one has
$B(r,v) \propto \log(v_\xi/v) + C \log(r)$.
Thus $B(r,v) \simeq r^2$ implies that $r \sim \log(v_\xi/v)^{1/2}$,
and therefore $B(r,v) \simeq B(0,v)$.
If one does so
and since $B(0,v) \gg 1$, one can restrict oneself to $K=0$ or $K=\pm K_0$
in  (\ref{offdiag}). The off diagonal part of the self energy
becomes
\begin{equation} \label{finalform}
\sigma(q,v) \simeq \lambda B(0,v)^{-1-\lambda/2} + c_1 q^2 e^{-\frac{q^2}2
B(0,v)}
              + c_2 e^{-\frac{K_0^2}2 B(0,v)}
\end{equation}
The two first terms in (\ref{finalform}) come from the $K=0$ part in
(\ref{offdiag}). The term in $\lambda$ is due to the extra unphysical
disorder and, as long as $\lambda$ is finite dominates the long range
behavior ($q\to 0$ and large $B(0,v)$). The other terms
correspond to the same separation of the Fourier components of the
random potential than in the discussion leading to our model (\ref{cardyos}).
The $q \simeq 0$ components gives the same contribution than the terms
in (\ref{cardyos}) in $\nabla u_a \nabla u_b$.
Indeed this $K=0$ term comes from
\begin{equation}
\langle \int dr dr' \delta(r - r' + u^a(r) - u^b(r')) \rangle
\end{equation}
This term measures the smooth change in local density due to
the slowly varying displacement field, and keeping only this term
amounts to neglect the discreteness of the vortices in the original
Hamiltonian.
The third term in (\ref{finalform}) comes from the $q \simeq K_0$
component of the disorder, i.e. the part which has
the periodicity of the equilibrium lattice. For the physical disorder
$\lambda = 0$ (\ref{finalform}) is
identical to the one obtained from the local model (\ref{cardyos}).

In fact neglecting the $x$ dependence in $B(x,v)$ is equivalent to
the replacement $\phi(x) = x - u(x)$, i.e identifying
$u(\phi(x))-u(\phi(y))$ with $u(x)-u(y)$,
which led to the model (\ref{cardyos}).
Such a replacement becomes exact in the elastic limit
$\overline{\langle (\nabla u)^2 \rangle} \ll 1$ which has been checked
self consistently on our solution.

\section{Random manifold in $d=2$} \label{random2}

Let us consider the original model (\ref{cardyos}) in $d=2$
keeping all the harmonics. We will treat only the simpler
case of isotropic elastic matrix, but for arbitrary $m$ and
lattice symmetry.
Similarly to the study in $d > 2$ we use the rescaled quantities
(\ref{dimensionless}) which satisfy the rescaled equations
(\ref{implicit}). In $d=2$ one has $\xi=a^2 c \sqrt{m/4 \pi^3 \Delta_{K_0}}$
and $v_\xi=\pi T/a^2 c$. The equations read:
\begin{eqnarray}
[s](y) & = &  h(z) \nonumber \\
y & = & -  h'(z)/h(z) \label{minimum}
\end{eqnarray}
where $h(z)$ is defined in (\ref{deff}). It appears clearly on (\ref{minimum})
that $y$ has a minimum value which is
\begin{equation}
y_0 =  \left(\frac{K_0 a}{2 \pi}\right)^2
\end{equation}
Thus the function $[\sigma](v)$ must vanish
for $v < v_0=v_\xi y_0=T/T_c$ which is identical to the value of
the breakpoint $v_c$ in (\ref{vc}) of the one step solution
for the single cosine model.
The asymptotic behavior at large distance will
thus be identical to the one of the single cosine model studied in
section~\ref{singlecos}. The transition temperature also turns out to be
exactly the same $T_c=4 \pi c/{K_0}^2$.
However, for $y>y_0$, $[s](y)$ will increase continuously until
the breakpoint $y=y_c$ above which it becomes constant again.
Using equation (\ref{nonuniversal}) the breakpoint $z_c = b(y_c)$ satisfies
\begin{equation} \label{equabas}
z_c = \frac{\pi T}{c a^2} \log[1+ \frac{\Lambda^2 \xi^2}{[s](y_c)}]
\end{equation}
where we used the circular cutoff $\Lambda = 2\pi/a$.
Using the equation (\ref{minimum}) one sees that $z_c$ is determined
as the solution of the equation $z_c=f(z_c)$ where we have defined the
function $f(z)$ as:
\begin{equation}
f(z) = \frac{\pi T}{c a^2} \log[1+ \frac{\Lambda^2 \xi^2}{h(z)}]
\end{equation}

Let us study the function $f(z)$. Using the two limiting forms
for $h(z)$, i.e
\begin{eqnarray}
h(z) & = & A  e^{- \alpha z} ~~~ z >>1  \\
h(z) & = & B/z^{(m+4)/2}  ~~~ z <<1
\end{eqnarray}
from \ref{largez} and \ref{smallz}, one finds that the
function $f(z)$ has three different behaviours depending on the values
of $z$
\begin{eqnarray}
f(z) &\simeq& \frac{\pi T}{c a^2} ( \alpha z + \log((\xi/a)^2/A)  )
\qquad z \gg 1 \nonumber \\
f(z) &\simeq& \frac{\pi T}{c a^2} ( \frac{m+4}{2} \log(z) + \log((\xi/a)^2/B)
)
\qquad  (a/\xi)^{4/(m+4)} \ll z \ll 1 \label{regimes} \\
f(z) &\simeq& \frac{\pi T}{B c a^2} (\xi/a)^2 z^{(m+4)/2}
\qquad  z \ll (a/\xi)^{4/(m+4)} \nonumber
\end{eqnarray}
taking into account that $\xi/a >>1$.

The first regime corresponds to high temperatures. One finds
no solution when $T > T_c$ and when $T<T_c$ one finds
$z_c= \pi T \log((\xi/a)^2/A) / c a^2 (1-T/T_c)$.
In that regime $y_c$ and $y_0$ are
very close to each other and the solution is very similar to
the one step solution of the single cosine model. There
is no true random manifold regime.

Solving the equation in the second regime one finds
approximately
\begin{equation}
z_c \simeq \frac{\pi T}{c a^2} \log((\xi/a)^2/B)
{}~~~~ v_c \simeq \ \frac{m+4}{2 \log((\xi/a)^2/B)}
\end{equation}
where we have used that in that regime $y \simeq (m+4)/(2 z)$.
The condition $z_c \ll 1$ shows that this regime exists only
at low temperature:
\begin{equation}
T < T^* = \frac{c a^2}{\pi \log((\xi/a)^2/B)}
\end{equation}
There all the
harmonics contribute, $y_c$ and $y_0$ are very different from
each other and there is a large random manifold regime, between the Larkin
regime and the asymptotic regime. In this regime one has $\tilde{B}(x)
\sim A (x/\xi)^{2 \nu}$ with $\nu=\nu_{RM}=2/(4+m)$.

It is easy to see that the last regime of behaviour of $f(z)$ never
contributes.
This is because one is restricted to
values of $z$ such that $v_c=y_c v_\xi < 1$ which corresponds to
$z_c > c a^2 / (T \pi)$.
This regime where there is no breakpoint ($l \sim a$),
and therefore no Larkin regime, as already discussed after
formula (\ref{nobreakpoint}),
arises at very low temperatures $T<
T'^* \sim (c a^2/\pi) (a/\xi)^{4/(4+m)}/\log((\xi/a)^2/B)$


\begin{thebibliography}{10}

\bibitem{junk}
Laboratoire associ\'e au CNRS. email: giam@lps.u-psud.fr.

\bibitem{frad}
Laboratoire Propre du CNRS, associ\'e a l'Ecole Normale Sup\'erieure et a
  l'Universit\'e Paris-Sud. email: ledou@physique.ens.fr.

\bibitem{feigelman_collective}
M. Feigelman, V. Geshkenbein, A. Larkin, and V. Vinokur, Phys. Rev. Lett. {\bf
  63},  2303  (1989).

\bibitem{fisher_vortexglass_short}
M.~P.~A. Fisher, Phys. Rev. Lett. {\bf 62},  1415  (1989).

\bibitem{nelson_columnar_long}
D.~R. Nelson and V.~M. Vinokur, Phys. Rev. B {\bf 48},  13060  (1993).

\bibitem{hwa_splay_prl}
T. Hwa, P.~L. Doussal, D.~R. Nelson, and V.~M. Vinokur, Phys. Rev. Lett. {\bf
  71},  3545  (1993).

\bibitem{revue_russes}
For a very recent review see G. Blatter, M. V. Feigel'man, V. B. Geshkenbein,
  A. I Larkin and V. M. Vinokur, to be published.

\bibitem{gruner_revue_cdw}
G. Gr{\" u}ner, Rev. Mod. Phys. {\bf 60},  1129  (1988).

\bibitem{wigner_andrei}
E.~Y.~A. et~al., Phys. Rev. Lett. {\bf 60},  2765  (1988).

\bibitem{wigner_jiang}
H.~W.~J. et~al., Phys. Rev. Lett. {\bf 65},  633  (1990).

\bibitem{wigner_shklovskii}
I.~M. Rusin, S. Marianer, and B.~I. Shklovskii, Phys. Rev. B {\bf 46},  3999
  (1992).

\bibitem{wigner_millis}
B.~G.~A. Normand, P.~B. Littlewood, and A.~J. Millis, Phys. Rev. B {\bf 46},
  3920  (1992).

\bibitem{seshadri_bubbles_thermal}
R. Seshadri and R.~M. Westervelt, Phys. Rev. B {\bf 46},  5142  (1992).

\bibitem{seshadri_bubbles_long}
R. Seshadri and R.~M. Westervelt, Phys. Rev. B {\bf 46},  5150  (1992).

\bibitem{vinokur_josephson_short}
V.~M. Vinokur and A.~E. Koshelev, Sov. Phys. JETP {\bf 70},  547  (1990).

\bibitem{balents_josephson_long}
L. Balents and S. H. Simon, Preprint ITP 1994.

\bibitem{toner_log_2}
J. Toner and D. DiVincenzo, Phys. Rev. B {\bf 41},  632  (1990).

\bibitem{pokrovsky_talapov_prl}
V.~L. Pokrovsky and A.~L. Talapov, Phys. Rev. Lett. {\bf 42},  65  (1979).

\bibitem{nelson_halperin_melting}
D.~R. Nelson and B.~I. Halperin, Phys. Rev. B {\bf 19},  2457  (1979).

\bibitem{fisher_vortexglass_long}
D.~S. Fisher, M.~P.~A. Fisher, and D.~A. Huse, Phys. Rev. B {\bf 43},  130
  (1990).

\bibitem{bokil_young_vglass}
Bokil and A.P. Young, {\it Absence of a Vortex Glass Transition in a
  Three-Dimensional Vortex Glass Model with Screening} , Preprint 1994.

\bibitem{giamarchi_columnar_variat}
T. Giamarchi and P. Le Doussal, in preparation.

\bibitem{larkin_70}
A. Larkin, Sov. Phys. JETP {\bf 31},  784  (1970).

\bibitem{larkin_ovchinnikov_pinning}
A.~I. Larkin and Y.~N. Ovchinokov, J. Low Temp. Phys {\bf 34},  409  (1979).

\bibitem{bouchaud_variational_vortex}
J. Bouchaud, M. M{\'e}zard, and J. Yedidia, Phys. Rev. B {\bf 46},  14686
  (1992).

\bibitem{bouchaud_variational_vortex_prl}
J. Bouchaud, M. M{\'e}zard, and J. Yedidia, Phys. Rev. Lett. {\bf 67},  3840
  (1991).

\bibitem{nattermann_pinning}
T. Nattermann, Phys. Rev. Lett. {\bf 64},  2454  (1990).

\bibitem{giamarchi_vortex_short}
T. Giamarchi and P. {Le Doussal}, Phys. Rev. Lett. {\bf 72},  1530  (1994).

\bibitem{mezard_variational_replica}
M. Mezard and G. Parisi, J. de Phys. I {\bf 4},  809  (1991).

\bibitem{chudnovsky_pinning}
E. Chudnovsky, Phys. Rev. Lett. {\bf 65},  3060  (1990).

\bibitem{grier_decoration_manips}
D. Grier {\it et~al.}, Phys. Rev. Lett. {\bf 66},  2270  (1991).

\bibitem{yaron_neutrons_vortex}
U. Yaron et al. ``Neutron Diffraction studies of flowing and Pinned Magnetic
  Flux Lattices in 2H-NbSe$_2$'' submitted to Physical Review Letter (1994).

\bibitem{edwards_replica}
S.~F. Edwards and P.~W. Anderson, J. Phys. F {\bf 5},  965  (1975).

\bibitem{haldane_bosons}
F.~D.~M. Haldane, Phys. Rev. Lett. {\bf 47},  1840  (1981).

\bibitem{fukuyama_pinning}
H. Fukuyama and P. Lee, Phys. Rev. B {\bf 17},  535  (1978).

\bibitem{korshunov_variational_short}
S.~E. Korshunov, Phys. Rev. B {\bf 48},  3969  (1993).

\bibitem{kwok_electron_defects}
W.~K. Kwok {\it et~al.}, Physica B {\bf 197},  579  (1994).

\bibitem{Huse_private}
D. A. Huse, private communication.

\bibitem{glazman_koshelev_decoupling}
L.~I. Glazman and A.~E. Koshelev, Phys. Rev. B {\bf 43},  2835  (1991).

\bibitem{fisher_c44_calcul}
D. S. Fisher, in {\it Phenomenology and Applications of High Temperature
  Superconductors}, edited by K. Bedell, M. Inui, D. Metzel,J.R. Schrieffer,
  and S. Doniach (Addison-Wesley, New York, 1991).

\bibitem{decoration_giamarchi_pld}
T. Giamarchi, P. Le Doussal, C.A. Murray, in preparation.

\bibitem{pld_machta_saw}
P. Le Doussal and J. Machta, J. Stat. Phys. {\bf 64} 541 (1991).

\bibitem{halpin_frg}
T. Halpin-Healy, Phys. Rev. A {\bf 42},  711  (1990).

\bibitem{marchetti_nelson_fits}
M.~C. Marchetti and D.~R. Nelson, Phys. Rev. B {\bf 47},  12214  (1993).

\bibitem{ledoussal_rsb_lettre}
P. Le Doussal and T. Giamarchi ``Replica symmetry breaking in the 2D XY model
  in a random magnetic field'' to appear in Phys. Rev. Lett. {\bf 74} January
  1995.

\bibitem{fisher_functional_rg}
D. Fisher, Phys. Rev. Lett. {\bf 56},  1964  (1986).

\bibitem{balents_frg_largen}
L. Balents and D.~S. Fisher, Phys. Rev. B {\bf 48},  5959  (1993).

\bibitem{natterman_leschhorn}
T. Nattermann and H. Leschhorn, Europhys. Lett. {\bf 14},  603  (1991).

\bibitem{villain_cosine_realrg}
J. Villain and J. Fernandez, Z Phys. B {\bf 54},  139  (1984).

\bibitem{cardy_desordre_rg}
J. Cardy and S. Ostlund, Phys. Rev. B {\bf 25},  6899  (1982).

\bibitem{batrouni_numerical_cardy}
G.~G. Batrouni and T. Hwa, Phys. Rev. Lett. {\bf 72},  4133  (1994).

\bibitem{cule_numerical_cardy}
D. Cule and Y. Shapir, Phys. Rev. Lett. {\bf 74},  114  (1995).

\bibitem{nattermann_flux_creep}
T.Nattermann, I.Lyuksyutov, and M.Schwartz, Europhys. Lett. {\bf 16},  295
  (1991).

\bibitem{nelson_elastic_disorder}
D.~R. Nelson, Phys. Rev. B {\bf 27},  2902  (1983).

\bibitem{seshadri_bubbles}
R. Seshadri and R. Westervelt, Phys. Rev. Lett. {\bf 66},  2774  (1991).

\bibitem{murray_colloid_hexatic}
C.~A. Murray, D.~H.~V. Winkle, and R. Wenk, Phase Transitions {\bf 21},  93
  (1990).

\bibitem{murray_colloid_prb}
C.~A. Murray, W.~O. Sprenger, and R. Wenk, Phys. Rev. B {\bf 42},  688  (1990).

\bibitem{safar_tricritical_prl}
H. Safar {\it et~al.}, Phys. Rev. Lett. {\bf 70},  3800  (1993).

\bibitem{charalambous_melting_rc}
M. Charalambous, J. Chaussy, and P. Lejay, Phys. Rev. B {\bf 45},  5091
  (1992).

\bibitem{chudnovsky_pinning_long}
E.~M. Chudnovsky, Phys. Rev. B {\bf 40},  11355  (1989).

\bibitem{kurchan_dynamics}
L.F. Cugliandolo and J. Kurchan, J. Phys A {\bf 27} 5749 (1994).

\end{thebibliography}

\end{document}